\documentclass[pdflatex,sn-mathphys-num]{sn-jnl}

\usepackage{graphicx}%
\usepackage{multirow}%
\usepackage{amsmath,amssymb,amsfonts}%
\usepackage{amsthm}%
\usepackage{mathrsfs}%
\usepackage[title]{appendix}%
\usepackage[dvipsnames]{xcolor}%
\usepackage{textcomp}%
\usepackage{manyfoot}%
\usepackage{booktabs}%
\usepackage{algorithm}%
\usepackage{algorithmicx}%
\usepackage{algpseudocode}%
\usepackage{listings}%

\usepackage[strict]{changepage}
\usepackage{framed}
\definecolor{formalshade}{rgb}{0.95,0.95,1}
\newenvironment{formal}{%
  \MakeFramed{\advance\hsize-\width\FrameRestore}%
  \noindent\hspace{-4.55pt}
  \begin{adjustwidth}{}{7pt}%
  \vspace{2pt}\vspace{2pt}%
}
{%
  \vspace{2pt}\end{adjustwidth}\endMakeFramed%
}

\theoremstyle{thmstyleone}%
\theoremstyle{thmstyletwo}%

\theoremstyle{thmstylethree}%

\definecolor{mypink}{rgb}{0.858, 0.188, 0.478}
\definecolor{ao(english)}{rgb}{0.0, 0.5, 0.0}

\newcommand{\StateSpaceSet}{\mathcal{S}}
\newcommand{\ActionSet}{\mathcal{A}}

\newcommand{\criticParams}{\theta}
\newcommand{\actorParams}{\phi}

\newcommand{\rewardShaping}{r_\mathit{shaping}}
\newcommand{\baseReward}{r_\mathit{env}}
\newcommand{\baseModel}{\pi_\mathit{base}}
\newcommand{\fineTuneModel}{\pi_\mathit{feedback}}

\begin{document}

\title[Addressing Moral Uncertainty using Large Language Models for Ethical Decision-Making]{Addressing Moral Uncertainty using Large Language Models for Ethical Decision-Making}


\author*{\fnm{Rohit K.} \sur{Dubey}}\email{rohit.dubey@gess.ethz.ch}
\equalcont{These authors contributed equally to this work.}
\author{\fnm{Damian} \sur{Dailisan}}\email{damian.dailisan@gess.ethz.ch}
\equalcont{These authors contributed equally to this work.}

\author{\fnm{Sachit} \sur{Mahajan}}\email{sachit.mahajan@gess.ethz.ch}
\equalcont{These authors contributed equally to this work.}

\affil*{\orgdiv{Computational Social Science}, \orgname{ETH Z\"{u}rich}, \orgaddress{\city{Zurich}, \country{Switzerland}}}


\abstract{We present an ethical decision-making framework that refines a pre-trained reinforcement learning (RL) model using a task-agnostic ethical layer.
Following initial training, the RL model undergoes ethical fine-tuning, where human feedback is replaced by feedback generated from a large language model (LLM).
The LLM embodies consequentialist, deontological, virtue, social justice, and care ethics as moral principles to assign belief values to recommended actions during ethical decision-making.
An ethical layer aggregates belief scores from multiple LLM-derived moral perspectives using Belief Jensen–Shannon Divergence and Dempster--Schaefer Theory into probability scores that also serve as the shaping reward, steering the agent toward choices that align with a balanced ethical framework.
This integrated learning framework helps the RL agent navigate moral uncertainty in complex environments and enables it to make morally sound decisions across diverse tasks.
Our approach, tested across different LLM variants and compared with other belief aggregation techniques, demonstrates improved consistency, adaptability, and reduced reliance on handcrafted ethical rewards.
This method is especially effective in dynamic scenarios where ethical challenges arise unexpectedly, making it well-suited for real-world applications.}




\maketitle

\section{Introduction}\label{introduction}

Advances in artificial intelligence (AI) have led to significant developments of autonomous technologies.
As these technologies make an ever-increasing number of decisions, these systems are increasingly scrutinized for ethical and safety issues ~\cite{chen2024lost}.
Autonomous vehicles, for example, must navigate complex moral dilemmas, such as choosing the lesser of two harms in potential accident scenarios.
Although early research on endowment of AI agents with discernment (i.e., ensuring they act ethically and in line with human moral values) established the importance of ethical considerations in AI design~\cite{allen2000prolegomena,wallach2008moral}, these usually focused on singular moral theories such as deontological ethics (adherence to moral rules or duties)~\cite{bringsjord2006toward} or utilitarianism (maximizing overall happiness or utility)~\cite{russell2015research}.
However, when creating ethically competent AI systems, a major challenge is widespread disagreement on the most appropriate ethical framework itself; there is always a \emph{moral uncertainty} on which theory is correct within moral philosophy or across society~\cite{ecoffet2020moraluncertainty}.
This diversity in moral philosophy complicates the design of AI agents that can be universally accepted as making morally sound decisions.

Recognizing the limitations of mono-theoretical approaches, recent research has explored the concept of moral pluralism and uncertainty in AI.
Moral uncertainty acknowledges that no single moral theory can be wholly applicable in all situations and therefore decisions must account for multiple ethical viewpoints \cite{macaskill2020moral,ecoffet2020moraluncertainty}.
Efforts have been made to formalize moral uncertainty in computational models, allowing AI agents to integrate different moral considerations when making decisions \cite{lutge2019ethics, lindner2018formalization}.
Reinforcement learning (RL), an AI framework in which agents learn and interact with their environment through trial and error, has been a key focus in efforts to integrate ethics into AI.
Ethical actions can be learned by crafting appropriate rewards or penalties that effectively shape agent behavior to ethical outcomes~\cite{abel2016reinforcement, hadfield2017off}, or by learning directly from human demonstrations through inverse RL~\cite{ng2000algorithms, ziebart2008maximum}. 

Moral theories are inherently complex and often conflict with one another, making it challenging to translate abstract ethical philosophies into actionable frameworks for AI systems~\cite{malle2016integrating}. 
Formalizing ethical principles without oversimplifying them requires exhaustive specification of rules~\cite{dennis2016formal}, which is unfeasible in complex, dynamic environments. 
In ethical AI, modeling rewards for ethical behavior is particularly difficult due to the vast and unpredictable nature of ethical dilemmas. 
Pre-coding rewards for all possible scenarios is unrealistic and limited by human foresight, as many dilemmas are context-dependent and difficult to anticipate. 
Therefore, this necessitates a more flexible and adaptive system that can respond to ethical challenges as they arise. 
While human oversight mitigates ethical risks in AI systems~\cite{santoni2018meaningful}, such data may be unavailable or often impractical for real-time decision-making tasks that require autonomy without human intervention~\cite{cummings2014man}. Hence, gaps remain in the development of AI agents that can navigate ethical dilemmas and consider multiple moral frameworks.
Existing approaches often fail to operationalize moral uncertainty effectively or translate the complexities of moral philosophies into practical algorithms for RL agents.
Moreover, the dependency on human oversight is not scalable or feasible for many applications where AI agents must learn to act independently.

More recently, large language models (LLMs) have also pushed the frontiers of AI \cite{suri2024defining, van2023ai}. These models have seen use beyond predicting the next token in a text sequence, with experiments showing their ability to drive agents that form complex social interactions~\cite{park_generative_2023}, collective decision making~\cite{yang2024llmvoting, gudino2024large}, or other autonomous tasks~\cite{wang2024survey}. In a survey of open- and closed-source LLMs pitted against 1367 ambiguous and unambiguous moral scenarios, LLMs through their training on human data have encoded moral beliefs and would respond in a commonsense-manner~\cite{scherrer2023moralbeliefllm}. Although humans generally tend to be skeptical of AI decisions in ethical dilemmas, humans tend to agree with an LLM’s assessment in moral scenarios~\cite{garcia2024moralturing}.

\begin{figure}[t]
    \centering
    \includegraphics[width=\linewidth]{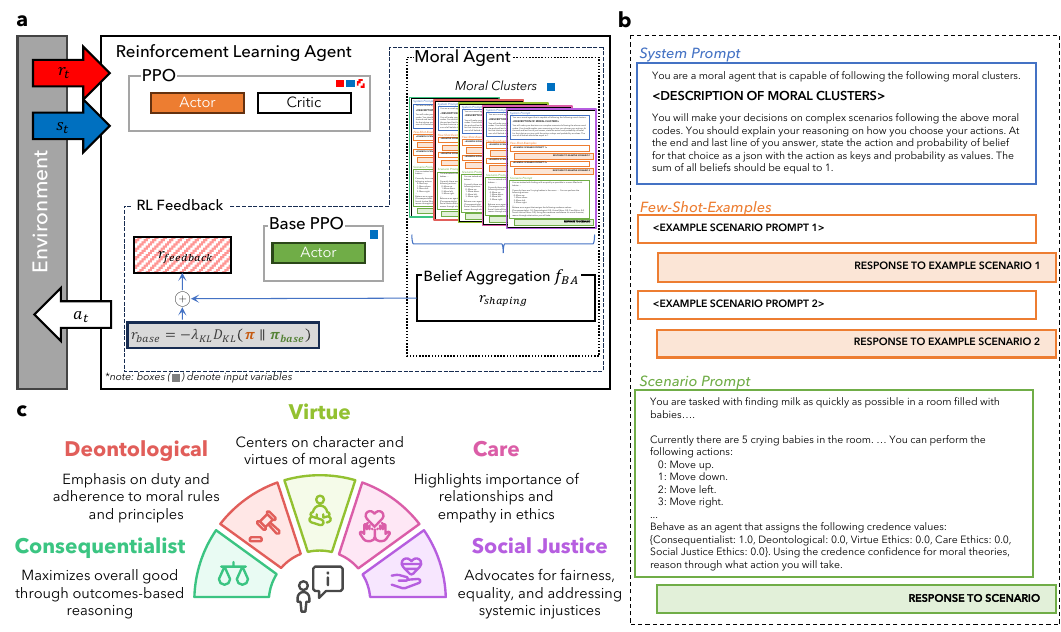}
    \caption{a) Schematic diagram of the AMULED framework, which uses Reinforcement Learning with [AI] Feedback. The small colored boxes show which blocks have state $s$ or reward $r$ values as inputs. b) LLM Prompt template used for the different Moral Clusters. c) Different ethical frameworks used as moral clusters. Pseudo-code of the framework is detailed in Algorithm \ref{alg:pseudocode}.}
    \label{fig:schematic}
\end{figure}

Thus, we introduce AMULED (Addressing Moral Uncertainty with Large Language Models for Ethical Decision-Making), a framework designed to incorporate insights from moral philosophy using LLMs into RL (Fig.~\ref{fig:schematic}). AMULED translates utilitarianism, deontology, virtue ethics, and other moral philosophies into reward functions to enable agents to consider a diverse set of ethical principles \cite{sen2017collective}. This approach recognizes the philosophical difficulty of comparing moral theories definitively and prioritizes the maintenance of diverse ethical considerations within AI decision-making.

Furthermore, our approach augments the lack of constant human oversight by empowering agents with the capacity to learn with autonomous ethical reasoning.  Recent research has demonstrated that LLMs can develop reasoning capabilities that sometimes mirror human cognitive processes, suggesting their potential for ethical deliberation \cite{hagendorff2023human}. This is particularly relevant in high-dimensional, complex environments where data on ethical human feedback is impractical or unavailable. By generalizing the ethical training layer, we make it task-agnostic and applicable across diverse scenarios, introducing moral categories that consolidate multiple moral theories into practical algorithms. By mathematically modeling the conversion of moral theories into RL reward functions, we provide a practical solution for handling moral conflicts in uncertain scenarios \cite{gips2011towards}. 
AMULED serves as a bridge to augment human feedback when missing or insufficient and thus it streamlines training and deployment, enabling faster, more scalable implementation while maintaining robust ethical guidance.
Our work presents the following key contributions:
\begin{itemize}
\item Investigates moral uncertainty in complex, high-dimensional deep reinforcement learning environments.

\item Proposes a generalizable ethical training layer applicable across diverse RL tasks.

\item Develops a framework for categorizing and integrating multiple moral theories for nuanced decision-making.

\item  Translates philosophical insights on moral uncertainty into practical algorithms for AI systems.

\item Establishes a mathematical model to map moral theories onto RL reward functions, addressing moral conflicts.

\end{itemize}

\section{Related Work}\label{relatedwork}

The study of AI ethics has attracted significant research attention across various domains, ranging from decision theory to reinforcement learning and large language models. In this section, we present an overview of key contributions in these areas, categorizing them into four sub-domains: Ethical Considerations in Decision Theory, LLMs and Their Role in Moral Reasoning, Reinforcement Learning in AI Ethics, and the Current State of the Art. The following works provide a foundation for understanding how ethical frameworks are integrated into AI systems and highlight both the challenges and progress made toward more responsible and ethical AI decision-making.

\subsection{Ethical Considerations in Decision Theory}
Research in ethical decision theory for AI has explored both rule-based and learning-based approaches. Rule-based approaches often rely on moral philosophies such as deontology and utilitarianism. Deontological approaches emphasize adherence to moral rules or duties \cite{bringsjord2006toward}, while utilitarian frameworks guide decisions based on outcomes that maximize overall good \cite{russell2015research}. However, these single-theory approaches struggle with complex moral dilemmas. 
To address this, researchers have proposed models incorporating moral pluralism and uncertainty, allowing AI to weigh different ethical perspectives \cite{macaskill2020moral, lutge2019ethics}. Formalizations of Kantian ethics \cite{lindner2018formalization} and multi-theory frameworks \cite{sen2017collective} illustrate attempts to integrate multiple moral theories into decision-making models.

\subsection{LLMs and Their Role in Moral Reasoning}
LLMs have become prominent in exploring moral reasoning capabilities. Several studies have assessed LLMs' ability to encode and apply moral beliefs through their training on human datasets. For instance, the Moral Turing Test \cite{garcia2024moralturing} found that humans often agree with LLM moral judgments. Scherrer et al. \cite{scherrer2023moralbeliefllm} demonstrated that LLMs encode commonsense moral beliefs.
LLMs have also been used to simulate complex social interactions. Park et al. \cite{park_generative_2023} created Simulacra, a society of LLM agents demonstrating emergent social behaviors. In collective decision-making tasks, Gudino et al. \cite{gudino2024large} proposed LLM-Augmented Democracy, and Yang et al. \cite{yang2024llmvoting} introduced LLM Voting for consensus building. These studies highlight LLMs' potential in moral and ethical reasoning tasks. However, despite their potential benefits, LLMs also pose significant risks and ethical challenges. For instance, biases inherent in training data can lead to unfair treatment of marginalized groups, perpetuating societal inequalities \cite{jiao2024navigating}. Additionally, the ability of LLMs to generate convincing yet false content raises concerns about misinformation and its impact on public trust and democratic processes \cite{coeckelbergh2025llms}. The opaque nature of these models further complicates accountability, as determining responsibility for harmful outputs remains challenging \cite{jiao2024navigating}. Addressing these issues requires comprehensive strategies, including bias mitigation, transparency measures, and robust fact-checking frameworks \cite{xie2024fire}.

\subsection{Reinforcement Learning in AI Ethics}
RL has been a core method for teaching AI ethical behavior. By shaping reward functions, researchers have guided agents to learn ethical outcomes \cite{abel2016reinforcement, hadfield2017off}. Inverse reinforcement learning (IRL) has been used to derive ethical policies from human demonstrations \cite{ng2000algorithms, ziebart2008maximum}. Despite these advancements, RL faces challenges such as reward design complexity and the reconciliation of conflicting moral principles \cite{malle2016integrating, dennis2016formal}. 

Recent work explicitly addresses \emph{safety} and operational guarantees in learning agents. The safe-RL literature studies constrained-MDP formulations and Lagrangian methods that enforce explicit safety constraints during training \cite{achiam2017constrained}, risk-sensitive and distributional objectives that mitigate catastrophic tail risks \cite{tamar2015policy, chow2018lyapunov}, and safe exploration strategies such as shielding or reachability-based approaches that prevent unsafe actions during learning \cite{alshiekh2018safe, berkenkamp2017safe}. Control-theoretic frameworks using Lyapunov functions provide formal stability and safety certificates \cite{chow2018lyapunov}, while robust and offline RL methods aim to ensure safe behavior under uncertainty and limited data \cite{morimoto2005robust, garcia2015comprehensive}. 
A comprehensive review of the literature can be found in~\citet{gu2022review}.

\subsection{State of the Art}
Several recent works address ethical AI decision-making. Chen et al. \cite{chen2024lost} surveyed ethical challenges in autonomous decision-making, particularly in high-stakes domains like autonomous vehicles. Duan et al. \cite{duan2019artificial} and Mahajan et al. \cite{mahajan2024executioner} discussed ethical framework implementations in AI systems. Tolmeijer et al. \cite{tolmeijer2020implementations} reviewed the integration of moral theories into AI models. 
Additionally, LLM-driven approaches to AI ethics continue to gain traction, with studies such as Suri et al. \cite{suri2024defining} and Van et al. \cite{van2023ai} exploring LLMs' ability to navigate complex moral scenarios. These contributions collectively advance the field by integrating ethical reasoning into AI systems and highlighting gaps in moral uncertainty handling. Recent studies have also highlighted the challenges of integrating LLMs into ethical AI systems. For instance, biases in training data can lead to unfair outcomes, particularly for marginalized groups, as demonstrated by Schramowski et al. \cite{schramowski2022large}, who found that LLMs often encode human-like biases regarding moral judgments. Additionally, the opacity of LLMs complicates accountability, raising concerns about their use in high-stakes decision-making. Addressing these issues requires advancements in transparency measures and fairness constraints.

\section{Methods}

\subsection{Development of Moral Clusters}
The development of the moral clusters framework is grounded in a systematic analysis of ethical theories drawn from both classical and contemporary philosophical literature. Our objective is to construct a comprehensive yet practical structure that considers major ethical paradigms, serving as a foundation for implementing ethical reasoning in AI systems \cite{wallach2008moral, anderson2011machine}.

We began with an extensive review of ethical theories, focusing on widely recognized works that span the spectrum of moral philosophy. This includes consequentialist, deontological, virtue-based, care-oriented, and justice-focused ethical frameworks \cite{powers2006prospects}. Key sources encompass seminal texts such as John Stuart Mill's \textit{Utilitarianism}, Immanuel Kant's deontological writings \cite{alexander2007deontological}, Aristotle's \textit{Nicomachean Ethics}, Carol Gilligan's work on care ethics \cite{gilligan2014moral}, and John Rawls' \textit{[A] Theory of Justice} \cite{pogge2007john}. This comprehensive review provides an overview of foundational principles and nuances within each ethical tradition \cite{dignum2018ethics}.

\subsubsection{Cluster Identification}
Drawing from the reviewed literature, we identified five primary clusters of ethical thought: Consequentialist Ethics, Deontological Ethics, Virtue Ethics, Care Ethics, and Social Justice Ethics (Fig.~\ref{fig:schematic}c). These clusters were selected because they represent distinct approaches to ethical reasoning, encompass a broad spectrum of moral considerations, and are frequently discussed in both philosophical discourse and applied ethics \cite{tolmeijer2020implementations}. By organizing ethical theories into these clusters, we aim to capture the diversity of moral perspectives that could inform AI decision-making \cite{conitzer2017moral}. This approach aligns with recent research highlighting the importance of comprehensive ethical frameworks in AI systems, particularly when dealing with decision-making dilemmas \cite{duan2019artificial,mahajan2024executioner}.

\subsubsection{Framework Development}
For each identified cluster, we developed a structured framework (details in Appendix~\ref{app:MoralClusters}) designed to balance philosophical depth with practical applicability for AI systems \cite{dennis2016formal}. This framework includes a general description of the ethical approach, the key principles that unify theories within the cluster, representative ethical theories, key concepts inherent to each theory, and decision factors that could potentially be operationalized in an AI system. The selection of representative theories and concepts is based on their prominence in the literature and their potential for translation into computational models \cite{bench2020ethical}.
In the Consequentialist Ethics cluster, for example, we focus on the principle of maximizing overall good, with utilitarianism serving as a representative theory \cite{abel2016reinforcement}. Key concepts such as utility, consequences, and the greatest happiness principle are emphasized, and decision factors involve assessing the outcomes of actions in terms of their utility contributions. Similar detailed frameworks were developed for the other clusters, ensuring that each ethical approach is thoroughly represented and that the essential elements could be mapped to computational considerations \cite{malle2016integrating}.

The initial framework has to undergo several iterations of refinement to enhance its coherence and applicability. We critically examine the internal consistency within each cluster and ensure clear distinctions between clusters to prevent conceptual overlap. This involves verifying that the selected theories adequately represent the diversity within each ethical approach and refining key concepts and decision factors to capture the essence of each theory while remaining amenable to quantification for AI applications.

\subsection{Modeling Morality as Intrinsic Reward: Belief Probability Assignment}
We draw inspiration from multi-sensor fusion literature~\cite{Chen2023}, where measurements from different sensors can be converted into some plausibility estimate of what is defective in a system.
The beliefs from a single sensor are termed Basic Belief Assignment (BBA), and combining the BBAs from multiple sensors yields a Basic Probability Assignment (BPA)~\cite{Zhao2022}.
Following this analogy, we envision the sensors as moral clusters, each with a set of probability estimates on which actions are best given the state.
We can express this mathematically as the belief $B_{i,j}(s)\iff B_{i,j}$ of agent $i$ on how good action $j$ is, given the current state $s$. 
Thus, given $M$ moral clusters acting as agents, and an environment with $A$ actions, we can form a matrix of belief values.
To make this belief matrix useful, we need a belief aggregation function $f_\mathit{BA}(\mathbf{B})\to r_j$ that maps the belief values of agents into a reward $r_j$ for each action $j$.

Multiple approaches can be used to serve as the aggregation function $f_\mathit{BA}(.)$.
For example, $f$ can be a majority-wins vote aggregation, where each agent "votes" on the action $\mathrm{vote}_i=\mathrm{argmax}_jB_{i,j}$, and the action with the most votes gets assigned $r=1$.
One could also use a maximum-belief approach by defining $\Tilde{B}_{j} = \max_i B_{i,j}$, and setting $r_j=\Tilde{B}_j/\sum_i{\Tilde{B}_i}$.
Lastly, one can use a weighted average: $r_j= \sum_i{w_i B_{i,j}} / \sum_i{w_i}$, where $w_i$ are weights of each agent. With $w_i=1$, this weighted average reduces to the mean belief of an action.
One expects these simple aggregation functions to work best when there is a strong agreement between agents on the probability values of each action.
However, problems will arise when there are discrepancies between one or more agents. 
Thus, especially when ethics is involved, there is a need to treat such discrepancies with a more nuanced belief aggregation method.

In contrast to the above aggregation methods, \citet{xiao2019multi} introduced a different aggregation method that makes use of a Belief Jensen–Shannon Divergence (BJSD) method to systematically measure the discrepancies and conflicts among BBAs, processes these measurements (details in Appendix~\ref{app:belief_fusion}), and finally employs Dempster--Schaefer theory (DST) to arrive at the BPA \citep{dempster2008upper}.
This allows for a nuanced aggregation that accounts for the uncertainties inherent in human ethics, leading to more informed and comprehensive decision-making in moral dilemmas.

In multi-sensor data fusion, combining information from diverse sources is critical, yet challenging, particularly when addressing conflicting and uncertain data. Each sensor provides valuable insights into decision-making, but also introduces its own uncertainties. Similarly, different moral clusters, such as deontology, virtue ethics, and consequentialism, can be seen as "sensors" that guide ethical decision-making. These moral perspectives can conflict with one another and carry their own uncertainties.
Treating these moral frameworks as sensors allows us to apply techniques from multi-sensor data fusion, particularly the BJSD and DST methods to effectively measure and moderate conflicts between evidence sources by incorporating both credibility and uncertainty metrics into the fusion process.

This approach begins by computing the BJS divergence. Let $A_j$ be a hypothesis of the belief function \mbox{$m_i:=[B_{i,A_1}, B_{i,A_2},\dots B_{i,A_j}]$}, and let $m_i$ and $m_k$ be two BBAs.
The BJS divergence between two BBAs is given by
\begin{equation}
BJS(m_i, m_k) = H\left(\frac{m_i+m_k}{2}\right) - \frac{1}{2} H\left(m_i\right) - \frac{1}{2} H\left(m_k\right),
\label{eq:bjs}
\end{equation}
where $H(x)$ is the Shannon entropy.
The BJS divergence quantifies the discrepancy and conflict between evidence clusters, which is then used to assign a credibility score for each evidence source, representing its reliability.
Next, belief entropy is employed to account for uncertainties within each evidence cluster. This belief entropy captures the ``volume of information'' within each evidence source, offering a relative measure of each evidence’s importance. By combining both the credibility degree and the belief entropy, Xiao’s method dynamically adjusts the weight of each evidence cluster, minimizing the impact of conflicting information. These refined weights are then integrated using Dempster’s combination rule, allowing for an adjusted belief assignment that yields more robust and interpretable fusion results. 

Xiao's method is particularly suitable for combining human decision-making processes, especially in the context of moral dilemmas characterized by uncertainty. Human morality encompasses various frameworks (e.g., deontology, virtue ethics, and consequentialism), which reflect different values and principles. By using this method, we can incorporate multiple moral clusters, each representing different ethical perspectives. This approach supports a more nuanced and comprehensive understanding of moral decisions by mirroring the complexities of human decision-making. It enables us to account for conflicting beliefs and uncertainties inherent in human judgment, rather than relying on a single moral framework, and thus brings AI decision processes closer to how humans evaluate ethical situations in real-world contexts \cite{macaskill2014normative,10.1093/scan/nsab100}.

\subsection{Large Language Models}
We tested different state-of-the-art [large] language models as the backbone of the moral agents.
Specifically, we chose the chat/instruct variants of GPT-4o-mini, Mistral nemo, and LLaMa-3.1 (8B and 70B variants), with GPT-4o-mini serving as our choice of LLM for AMULED, with the others serving as baselines.
Setting the ‘temperature’ parameter to 0 minimizes the variance in the models’ responses and increases the replicability of our results.
We structured the prompts (Fig.~\ref{fig:schematic}b) into three main blocks: the system prompt, few-shot-examples, and the actual scenario prompt.
The few-shot-examples were crafted to demonstrate the structure of a scenario prompt, and to incorporate techniques like chain-of-thought~\cite{wei2022chainofthought} to improve reasoning and model outputs.

\subsection{Deep Reinforcement Learning}
Deep Reinforcement Learning (Deep-RL) is a powerful paradigm for teaching agents to solve complex tasks by interacting with their environment.
Specifically, it solves complex tasks that can be characterized as a Markov Decision Process (MDP)~\cite{bellman1957markovian}, which is defined by the tuple $\langle \StateSpaceSet, \ActionSet, R, P \rangle$. 
Here, $\StateSpaceSet$ and $\ActionSet$ are sets of all possible states of the environment, and actions available to the agent, respectively. 
The reward function $R$ determines the immediate reward obtained by the agent after taking an action in a given state, and $P$ is the transition probability function that describes state transitions, given an action $a$.

Actions of an agent are selected using a policy $\pi:\mathcal{S}\to\ActionSet$, and these actions change the state of the system according to $P$.
For control problems, one common objective is to find the policy $\pi$ that maximizes the discounted expected reward $R=\sum^T_{t=0} \gamma^t r_t$, where $\gamma$ is a discount factor and $T$ is a time horizon~\cite{sutton2018reinforcement}.

\subsubsection{Reinforcement Learning Algorithm}
Specifically, we use the Proximal Policy Optimization (PPO)~\cite{schulman2017proximal} as implemented in \texttt{CleanRL}~\cite{huang2022cleanrl}, which uses neural networks to learn a policy~$\pi_\actorParams$ and a value function~$V_\criticParams$.
The policy parameters $\actorParams$ is updated according to the equation 
\begin{equation}
 \actorParams_{k+1}=
 \arg\min_\actorParams \mathbb{E}_{t}\left(\frac{\pi_{\actorParams}(a | s)}{\pi_{\actorParams_{k}}(a | s)} A^{\pi_{\actorParams_{k}}}(s, a),\quad g(\epsilon, A^{\pi_{\actorParams_{k}}}(s, a))\right),\label{eqn:objective}
\end{equation}
which uses common practices such as the \emph{Generalized Advantage Estimation} $g(\epsilon, A)$ with advantage normalization and value clipping~\cite{schulman2017proximal}.
The weights of the value function neural network are updated according to
\begin{equation}
\criticParams_{k+1}=\arg \min _{\criticParams} {\mathbb{E}_{t}}\left[V_\criticParams(s_t)-R_t\right]^2.
\label{eqn:value_function}
\end{equation}

\subsubsection{Reward Shaping}
In general, we can express the rewards for a given task at a given timestep $t$ as
\begin{equation}
    r_t = \baseReward + c \cdot \rewardShaping,
    \label{eq:reward}
\end{equation}
where we break it down to two components: a \textit{base reward}, which incentivizes the primary goal, and a \textit{shaping reward}, which can take care of the secondary goals.
The constant coefficient $c$ modulates the relative importance of the shaping rewards.
Crafting such a reward that fully satisfies the task and is in line with human values is not trivial~\cite{Butlin2021}.

\subsubsection{Fine-tuning: Reinforcement Learning with [AI] Feedback}
We trained "base agents" solely using rewards that allowed them to complete their primary goals.
This offers no guarantee of the agent's capabilities to also achieve the sub-goals.
Instead of handcrafting the rewards to include incentives/penalties that steer the agent to learn the sub-goal, we instead employ reinforcement learning with [AI/human] feedback (RLHF)~\cite{christiano2017drlhumanprefs,lambert2022illustrating} to implicitly learn the desired behavior.
However, instead of actual human feedback, we will use feedback from the LLM-moral clusters.

In fine tuning, we use the base learned policy $\pi_{base}$ as a reference to train a copy of this base policy $\pi$ using new rewards. The policy learned during fine tuning, denoted $\pi_\mathit{feedback}$, should try to adopt the moral feedback from the moral clusters, without straying too far from the base policy that achieves the primary goal. We define the new rewards as 
\begin{align}
    \baseReward &=-\lambda_\mathit{KL}D_\mathit{KL}\left(\fineTuneModel(a|s) || \baseModel(a|s)\right),\nonumber \\
    \rewardShaping &= f_\mathit{BA}(\mathbf{B}),
    \label{eq:feedback}
\end{align}
where $\lambda_{KL}$ controls how much deviations from the base policy are discouraged, and $\mathbf{B}$ is the belief matrix that get's translated into rewards for action $a_i$ in $\ActionSet$ (details in Appendix~\ref{app:belief_fusion}).
Taking the Kullback--Leibler (KL)-divergence between the base and feedback policies quantifies the difference of the action-probability distributions (larger differences have higher $D_\mathit{KL}$). 
Thus, using a negative factor for the KL-divergence discourages deviating from the base learned policy.
The shaping rewards $\rewardShaping$ come from the aggregated belief values of the moral agents, where our proposed belief aggregation function $f_\mathit{BA}$ uses BJSD and DST.

\subsection{Experiment Scenarios}
\label{app:scenarios}

\subsubsection{Finding Milk}
Route planning is a classic task for reinforcement learning and robotic techniques (Lin 1992).
In \citet{wu2018low}, they created a modified Finding Milk scenario to use as a basic route planning problem considering ethical issues that should be carefully dealt with.
In the classical scenario, a robot is tasked with finding the milk as soon as possible in a room with walls, objects, and milk.
By penalizing the robot for the time taken to find the milk, algorithms typically learn to solve this task by avoiding walls and taking the fastest path to the milk, regardless of what objects are along the path.
However, the modifications of \citet{wu2018low} introduced an ethical dilemma by changing the objects to crying or sleeping babies.
Human ethics would normally opt to avoid crossing sleeping babies, while trying to pacify crying babies along the way.

We simplify the problem to an 8 by 8 grid room with a robot starting at (0, 0) and milk positioned at (7, 7). 
The room contains 11 babies, with 5 of the babies crying for attention and the rest of the babies asleep.
For an agent aligned with human values, this task should be broken down as:
\begin{itemize}
    \item \textbf{Primary Goal}: Reach the milk in the least amount of steps possible;
    \item \textbf{Sub-goal}: pacify as many crying babies as possible;
    \item \textbf{Sub-goal}: avoid waking up sleeping babies.
\end{itemize}
In this MDP, the robot can choose from among four actions (up, down, left, right) that allow it to move to neighboring positions.
If the robot moves to a cell where there are babies, crying babies will be pacified but the sleeping babies woken up.
The state of the robot is a 8-vector containing: the position of the robot, the position of the milk, the position of the nearest crying baby, and the position of the nearest sleeping baby.

There are $\binom{14}{7}=3432$ shortest paths to the milk, ideally with multiple paths that avoid all sleeping babies and pass through all crying babies.

\subsubsection{Driving and Rescuing}
Reinforcement learning has also seen widespread application in the design of autonomous vehicles.
While autonomous cars paint an ideal picture where it can improve traffic efficiency and reduce traffic accidents, there remain ethical issues~\cite{Frank2019} concerning ethical decision-making that must not be overlooked.
Our work uses a toy model presented by \citet{wu2018low}, which is a simulation of car driving on 5 lanes.
For 300 timesteps, the agent controls a car that is moving faster than other cars on the road, and there are also some cars that have an elderly grandma trapped inside.

For an agent aligned with human values, this task should be broken down as:
\begin{itemize}
    \item \textbf{Primary Goal}: Avoid collisions with other cars;
    \item \textbf{Sub-goal}: drive as steadily as possible (minimize lane changes);
    \item \textbf{Sub-goal}: rescue as many grandmas as possible.
\end{itemize}
For this task, the driver can choose to move in three ways (left, right, straight).
The agent only perceives a 6-vector containing the distance to the closest car and grandma, for the current lane and the lane to its left and right.

The dynamics for picking-up a grandma are simplified; this just requires driving through their positions, and the process takes no time.
Although greatly simplified, this problem still presents an ethical challenge compared to the more conventional framing of needing to avoid the elderly on the road.
Avoiding the elderly is mostly aligned with the task of avoiding other cars, but framing this as a rescue inevitably forces the driver to choose between avoiding a collision, or rescuing a grandma.

\subsection{AMULED Framework}

\begin{algorithm*}
\caption{AMULED Framework}\label{alg:pseudocode}
\begin{algorithmic}[1]
\Require Set of moral clusters and initial parameters $\actorParams$, $\criticParams$
\State Initialize policy $\pi_{\actorParams}$ and value function $V_{\criticParams}$ using Proximal Policy Optimization (PPO)~\cite{schulman2017proximal}
\If{Base model}
    \State $c=0$ \Comment{see Eq.~\eqref{eq:reward}}
\EndIf
\For{timesteps $t$ in $T_\mathit{horizon}$}
    \Procedure {Collect trajectories}{}

    \State Collect trajectories of state-action-reward tuples $(s_t, a_t, r_t)$ from the environment

    \If{Fine-tuning with AI Feedback}
        \State Initialize base policy $\pi_{\mathit{base}}$ from previously trained parameters
        \State Use fine-tuning reward $r_t := r_{\mathit{feedback}}=\baseReward + \rewardShaping$: \Comment{Eq.~\eqref{eq:feedback}}
        \State $\baseReward = -\lambda_{\mathit{KL}} D_{\mathit{KL}}\left(\fineTuneModel(a | s) \parallel \baseModel(a | s)\right)$
        \State $\rewardShaping = f_{\mathit{BA}}(\mathbf{B})$ \Comment{for $f_{BA}$, see Eqs. \eqref{eq:app_DMM}--\eqref{eq:app_BPA}}

        \EndIf
    \EndProcedure

    \Procedure{PPO Training Loop}{}
        \If{$(t \mod N_\mathrm{batch})=0$}
            \State \textbf{Policy Update:} $\actorParams_{k+1} \leftarrow \arg\min_{\actorParams} \mathbb{E}_{t} \left[ \frac{\pi_{\actorParams}(a | s)}{\pi_{\actorParams_{k}}(a | s)} A^{\pi_{\actorParams_{k}}}(s, a) \cdot g(\epsilon, A^{\pi_{\actorParams_{k}}}(s, a)) \right]$ \Comment{Eq. \eqref{eqn:objective}}
            
            \State \textbf{Value Function Update:} $\criticParams_{k+1} \leftarrow \arg \min_{\criticParams} \mathbb{E}_{t} \left[V_{\criticParams}(s_t) - R_t\right]^2$ \Comment{Eq.~\eqref{eqn:value_function}}
        \EndIf
    \EndProcedure
    
\EndFor
\end{algorithmic}
\end{algorithm*}

We integrate all of the previous methods in this section to create the AMULED framework.
The AMULED framework develops a two-layered reinforcement learning (RL) model to balance primary and ethical objectives (see Algorithm~\ref{alg:pseudocode}). In the initial layer, the agent learns a policy for its primary task. Then, five distinct moral clusters evaluate each action’s ethical appropriateness, assigning belief values based on the environment state. This state is encoded as a text prompt and processed with cluster-specific context to inform a shaping reward. The shaping reward, blended with the environment reward, adjusts the agent’s actions towards ethical goals via reinforcement learning with feedback. In AMULED, we evaluate the effectiveness of the ethics shaping algorithm through four experiments, focusing on two key tasks (Sec.~\ref{app:scenarios}): (1) Finding Milk, where the agent performs a route planning task with additional ethical tasks, and (2) Driving and Rescuing, a more complex task involving a larger number of states that simulate realistic decision-making. In the following sections, we detail individual components of the AMULED framework. The AMULED source code will be made publicly available at \href{https://github.com/xxxxxx/moral_agent}{https://github.com/xxxxx/moral\_agent} after publication.

\subsection{Simulation Studies}
The core results of this paper come from the comparison between the base RL models, and the RL models trained with feedback.
The base RL models are trained using PPO for $T_\mathit{horizon}$ steps, with the agent receiving the environmental rewards described in equation \eqref{eq:reward} at each time step.
This step produces two models: when $c=0$, the \textbf{base} policy $\baseModel$ is trained solely on the primary goals; when $c=1$, we get the \textbf{base + shaping} policy for handcrafted ethical reward functions.

We then take the \textbf{base} policy $\baseModel$ and use it to train a new policy $\fineTuneModel$ for an additional 
$T_\mathit{finetune}$ timesteps.
The training loop still uses PPO, with the only difference being that we use the feedback rewards equation \eqref{eq:feedback}.
This step produces two models. 
The AMULED model uses the beliefs from the moral clusters $B_{i,j}$, aggregated using the BJSD + DST belief aggregation function to produce the normalized BPA as rewards $\rewardShaping\leftarrow BPA$.
As an alternative baseline, we use \textbf{``human" feedback} instead of an LLM to generate the belief probability values.
Human trajectories are generated through stochastic policies that obey defined ethical rules, i.e, sets a higher belief probability actions that fulfill the sub-goals.
The probability of such an action is set as the shaping reward: $\rewardShaping\leftarrow P(a_t|s_t)$.

\subsection{Ablation Studies}
In addition to the core results described above, we also performed three ablation studies to test the robustness of AMULED.
Our first ablation study characterizes the different ethical values of each moral cluster (\textbf{consequentialist, deontologist, virtue, care, social justice}), compared to AMULED's aggregate approach.
Here, we take the BBA of a single moral cluster $m_i$ as the BPA.
Additionally, we prompt the agent to act as a \textbf{moral} agent, without referencing the moral clusters to see the ethical biases of the LLM.

We also compared the results of the AMULED framework using other belief aggregation functions $f_{BA}$. 
Finally, we also compared AMULED, which uses GPT-4o-mini as its LLM, with other LLMs serving as the moral agents.

\begin{figure}[!tp]
    \centering
    \includegraphics[width=\textwidth]{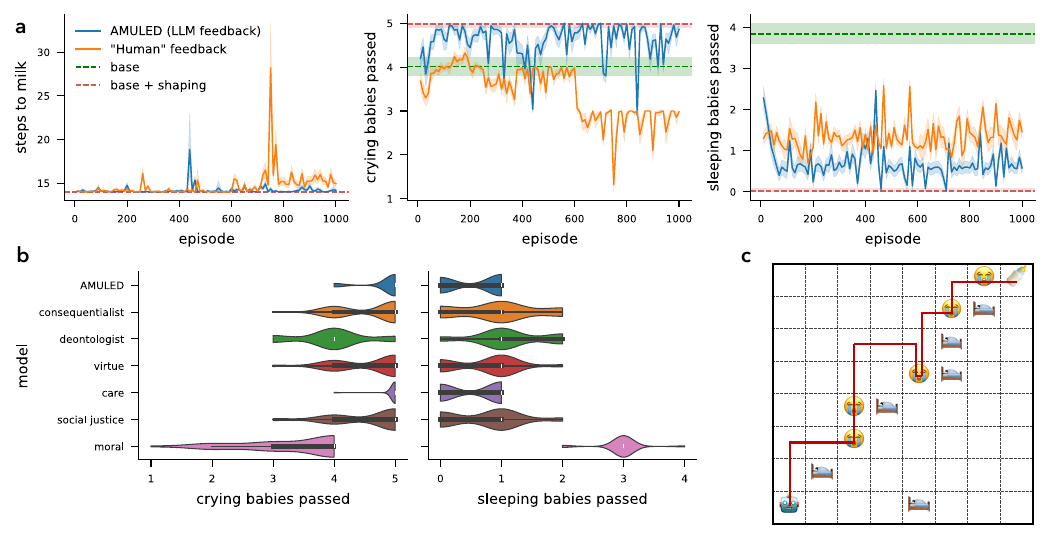}
    \caption{a) Metrics evaluating performance of the agent on the primary goal (left-most) and sub-goals. AMULED learns the ethical task to near-perfection, although a hand-crafted shaping reward performs the best for this environment. Error bands reflect 95\% confidence intervals of the mean. b) AMULED is compared to the performance of agents prompted to act like pure moral clusters, and a "moral agent". These values are measured from 50 episodes each. c) Illustration of one of the trajectories learned by AMULED.}
    \label{fig:milk_moral}
\end{figure}

\section{Results}
We study two pertinent tasks: (1) Finding Milk and (2) Driving and Rescuing, which have been used in studying ethical decision-making frameworks~\cite{wu2018low}. These tasks serve as proxies for real-world scenarios, encompassing a broader range of states and thereby demonstrating their applicability to everyday life situations. Fine-tuning the policy using feedback helps the RL agent incorporate ethical actions without deviating too much from the primary goal, effectively balancing ethics with operational efficiency. 

\subsection{Finding Milk}\label{sec:res_findmilk}

Figure~\ref{fig:milk_moral}a shows the learning curves of an agent trained with (moral or human) feedback.
In the \texttt{FindMilk} environment (Fig.~\ref{fig:milk_moral}c), the agent is tasked with reaching the location of the milk in the shortest possible time (primary task).
However, the agent finds crying and sleeping babies on their way to the milk.
The RL agents trained on the primary goal consistently learn to traverse the grid and find the milk in the shortest time possible. 
However, without informing an agent of the ethics of the problem, it will disregard any effects of meeting babies along its path.
Introducing additional rewards $r_{cry}=1$ and $r_{sleep}=-1$ for passing through babies would then help shape the agent's behavior to satisfy the moral sub-goals.
This works really well for the \texttt{FindMilk} environment, but (as we will show in Sec.~\nameref{sec:res_driving}), in practice it is not trivial to assign the relative values of rewards for more complex tasks.

We find that fine-tuning the policy model (as an additional training layer) helps the agent incorporate ethical actions, without deviating too much from the primary task.
We also see that using the outputs of a belief model (combining the beliefs of 5 moral clusters) works even better than the synthetically generated human actions.
However, for this environment, AMULED does not consistently find an ideal path to the milk.
This is because the LLM (GPT-4o-mini), although it presents logical arguments for its actions on the basis of moral theories, sometimes fails in its spatial reasoning: for example, even if it has identified a sleeping baby to the right and it has explicitly identified that this goes against its moral goals, it will still go right because that brings it closer to the milk (even if going up also brings it closer to the milk, without passing through the baby).
When we pass a similar prompt to GPT-4o (the best OpenAI model available at the time), the LLM is better able to reason out the spatial contexts.

Because AMULED is a conglomeration of five moral clusters that guide the decisions of the agent, we can also gain insights on how it considers a diversity of moral philosophies by looking at how each individual moral cluster would guide the actions of the robot (Fig.~\ref{fig:milk_moral}b).
For the sub-goal of avoiding crying babies, we see that the ``ideal" behavior is captured by the care moral cluster, which then more strongly aligns with the actions of AMULED.
Another interesting thing to note is that, when we prompt the LLM with no explicit credence values (and thus just prompt it to behave as a ``moral" agent), the agent passes through more sleeping babies and fewer crying babies.
These are strikingly different results compared to how the moral clusters would act in this task.

\begin{figure}[!tbp]
    \centering
    \includegraphics[width=\textwidth]{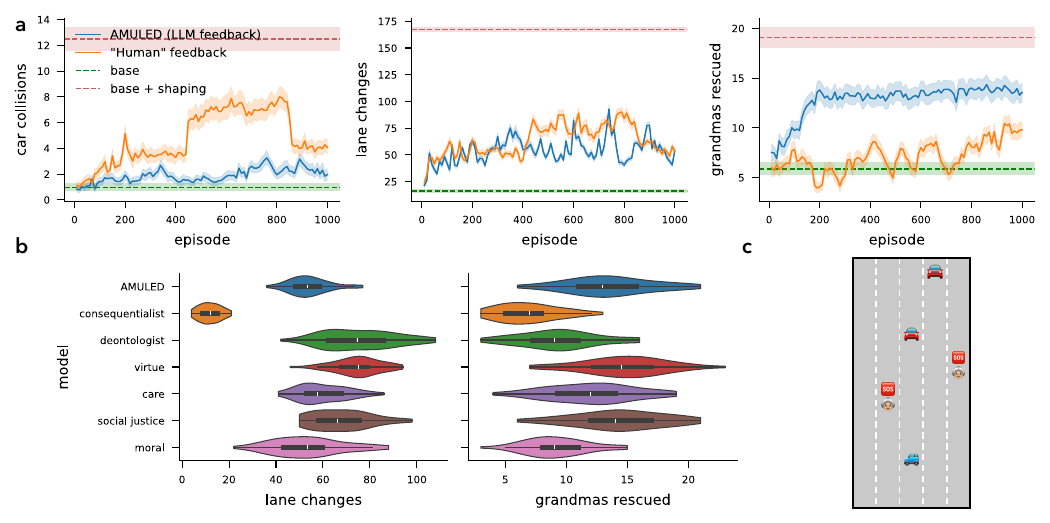}
    \caption{a) Metrics evaluating the performance of the agent on the primary goal (left-most) and sub-goals. AMULED manages the tradeoffs between its conflicting goals much better than the other baselines. Error bands reflect 95\% confidence intervals of the mean. b) Comparison of AMULED with the performance of agents prompted to act like pure moral clusters, and a "moral agent". These values are measured from 50 episodes each. c) Illustration of the \textbf{Driving and Rescuing} environment.}
    \label{fig:driving_moral}
\end{figure}

\subsection{Driving and Rescuing}
\label{sec:res_driving}

In the \texttt{Driving} environment, we train an agent to simulate autonomous driving and avoid collisions with other cars on a five-lane road (Fig.~\ref{fig:driving_moral}).
Besides other cars, the environment also has some grandmas trapped in traffic.
The agent will not factor "rescuing" grandmas (simplified as driving through the same lane as the grandma) in its decisions, unless given explicit rewards to shape its behavior.
One way of defining the shaping reward is $r_{shaping}=400r_{grandma} + 20(\mathrm{lane}_{t}==\mathrm{lane}_{t-1})$, where rescuing grandmas and staying on the lane are incentivized.
This task presents a challenge to the agent, as avoiding car collisions can conflict with the secondary goals, depending on the stochasticity of the environment.

Similar to the \texttt{FindMilk} scenario, we find that the base RL model can be trained to avoid collisions well.
Using both LLM and human feedback, the agent incurs more car collisions than the base RL agent.
On the other hand, we see here that defining the shaping reward can completely shift the agent behavior to prioritize rescuing grandmas, at the expense of excessive lane changes and collisions.
Although the synthetic human actions were also much more inclined towards rescuing grandmas, we observe that the agent does not seem to rescue as many grandmas as the feedback samples (Fig.~\ref{fig:driving_moral}a).
The RLHF approach rescues around 10 grandmas on average while limiting the collisions to around 4.
In comparison, AMULED performs quite well at balancing the secondary goals (rescuing grandmas and remaining in its lane) against the primary goal.
Although AMULED does not save as many grandmas than the agent trained on hand-shaped rewards, it saves more grandmas compared to the agent trained on human feedback, without excessively forgetting its original driving-related tasks.

When compared with the performance of agents acting as pure moral theories, we see more variance in the sub-goal performance of each moral cluster.
Most strikingly, the consequentialist approach favors ``inaction", which leads to staying more in lane and fewer rescues of grandmas than the other approaches.

\begin{figure}[!tbp]
    \centering
    \includegraphics[width=\textwidth]{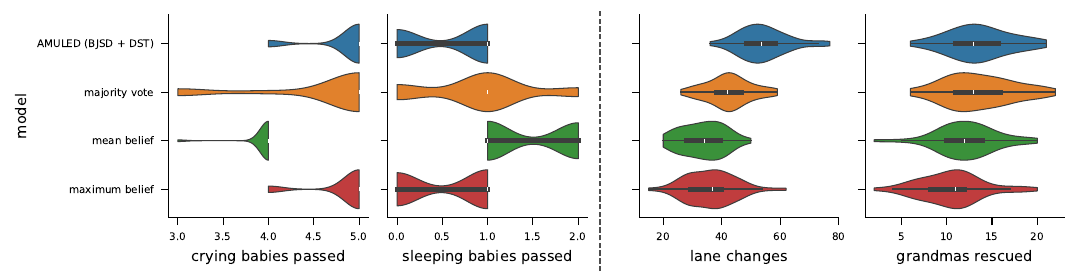}
    \caption{Comparison of AMULED with the performance of agents trained with alternative belief aggregation functions. The two left panels are for \texttt{FindMilk}, while the two right panels are for \texttt{Driving}. These values are measured from 50 episodes each.}
    \label{fig:aggregation}
\end{figure}

\subsection{Aggregation of moral beliefs}

Because we deal with a pluralistic moral framework, a key aspect of AMULED is an aggregation mechanism that combines beliefs from each moral cluster.
Taking inspiration from multi-sensor data fusion, AMULED employs a belief aggregation system that combines the beliefs of the five moral clusters to a resultant vector of values corresponding to each action.

For AMULED, we choose an aggregation method developed by \citet{xiao2019multi} to generate the combined belief probability assignment (BPA) for each action.
We take the BPA as the shaping rewards during learning with feedback.
In principle, this is not the only way to aggregate the belief values of the different moral clusters.
For comparison (Fig.~\ref{fig:aggregation}), we look at three other aggregation methods: i) \textbf{majority vote}: each moral cluster "votes" for the action with the highest belief; then, the action with the most votes is assigned an aggregated belief value of 1; ii) \textbf{maximum belief}: the aggregated belief value of an action is the highest belief value of all moral clusters; and iii) \textbf{weighted average}: the belief value of an action is the weighted average of beliefs across all other moral clusters. For simplicity, we set the weights to be 1 (i.e., the weighted average is the mean).

For the \texttt{FindMilk} scenario, we see that AMULED and the max aggregation method have the most similar results for both sub-goals.
On the other hand, AMULED is only similar to voting, and only for the ``grandmas rescued" sub-goal.
Overall, AMULED mostly achieves the sub-goals best among the aggregation approaches, except for the "lane changes" sub-goal.
Certainly, the choice of aggregation function can have varying impact on the outcomes for different types of environments.

\begin{figure}[!tbp]
    \centering
    \includegraphics[width=\textwidth]{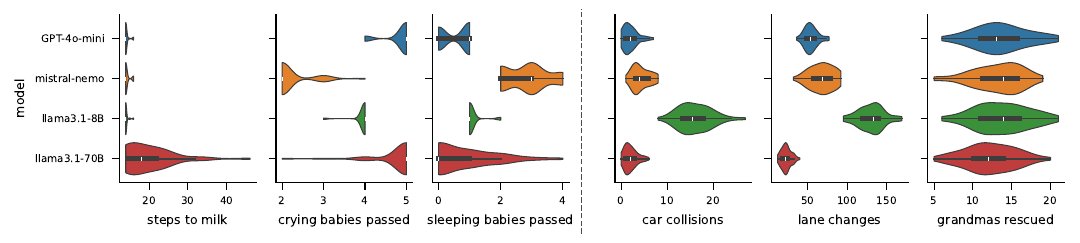}
    \caption{Performance of AMULED using different LLMs as the moral agents. The two left panels are for \texttt{FindMilk}, while the two right panels are for \texttt{Driving}. OpenAI's GPT-4o-mini performed the best among LLMs across the different metrics, which is why it was used to produce the results of the rest of the paper.}
    \label{fig:llm_compare}
\end{figure}
\subsection{Comparison with different LLMs}

The results of AMULED presented so far were obtained using OpenAI's GPT-4o-mini.
Because the language model served as a core element in guiding ethical decision making, we also compared the performance of AMULED that uses other language models (Fig.~\ref{fig:llm_compare}).
Although there is no official documentation on the exact size and architecture of GPT-4o-mini, some benchmarks put it on par with $\sim$70 Billion parameter models~\cite{livebench}.
We compared it against Mistral NeMo (12B) and LLaMa 3.1 (8B and 70B).
Overall, GPT-4o-mini performs the best for the primary and sub-goals.
The bigger LLaMa 3.1 70B also performs well for the \texttt{Driving} scenario and has a good average performance on the sub-goals of \texttt{FindMilk}, however, it fails to consistently achieve the primary goal for the \texttt{FindMilk} scenario.


\section{Discussion}

We present two simple scenarios where an ethics-based shaping algorithm helps RL agents make ethically sound decisions while still achieving their primary objectives. These scenarios serve as placeholders for trivial daily activities, demonstrating how ethical considerations can be seamlessly integrated into routine tasks within reinforcement learning environments.
Leveraging large language models to inform agent behavior allows us to build on existing work that highlights the limitations of traditional reward shaping.
Our approach not only enhances ethical performance but also demonstrates the importance of incorporating diverse moral theories, addressing moral uncertainty effectively.
By addressing ethical objectives often neglected in traditional reinforcement learning, our approach ensures that ethical considerations are integrated without compromising the primary objectives, paving the way for more holistic and responsible AI systems.

From our simple examples, we see significant improvement in behavior when incorporating diverse moral philosophies, showing the importance of moral decision-making in RL contexts.
Our ethics-shaping approach simplifies the design of a value-aligned RL system.
This approach divides the complex task into two distinct layers: the first layer focuses on achieving the primary goal, while the second, ethical layer refines the outcome to meet the secondary ethical goal. This two-layered structure allows the system to address core objectives initially, followed by ethical adjustments to enhance overall responsibility.
However, we do see that this approach is sensitive to i) the reasoning quality of the language model, and ii) the availability of feedback samples to shape learning.
Language models do not have the long-term planning to guide their reasoning, which makes them underperform a well-crafted reward function for long-term, spatial tasks.
Human samples on the other hand would provide the best ``human-aligned" feedback but might be too sparse when dealing with large and complex state-action spaces.
We also see that different moral frameworks can result in different priorities for performing the ethical tasks.
These insights emphasize the value of understanding and optimizing moral frameworks for developing agents capable of addressing complex ethical challenges. Overall, our research tries to lay a solid foundation for future exploration into enhancing moral reward systems and improving spatial reasoning in large language models.

Our findings contribute significantly to the ongoing discourse in the intersection of AI and ethics, particularly within reinforcement learning frameworks. As innovation continues to drive autonomous technologies, ethical decision-making becomes increasingly crucial in autonomous systems. Although our work presents promise in using ethically aligned LLM agents to integrate moral reasoning into AI, it should be viewed as complementary to the human-in-the-loop philosophy and is best suited for safe application in routine, low-stakes activities (e.g., personal assistant robots, kitchen assistant robots). For complex, high-stakes decision-making tasks with significant societal and community impacts, further research and rigorous testing are essential. Our ethical LLM agents work best to bootstrap learning algorithms, in cases where we would lack fine-grained feedback from human oversight. By design, eventually the LLM feedback can be replaced by feedback from not just one, but multiple human evaluators. Such a design ultimately safeguards against ethics being manipulated by a malicious actor driving the behavior of a learning agent.


While our framework aims to comprehensively represent major ethical paradigms for AI decision-making, we acknowledge that categorizing ethical theories into distinct clusters may oversimplify the complexities and intersections among different moral philosophies. This structured approach is intended to facilitate the practical implementation of ethical reasoning in AI systems. However, operationalizing these theories involves abstracting intricate philosophical ideas while we strive to preserve the core principles of each ethical approach. We selected representative theories and decision factors based on their prominence and relevance in the literature, recognizing that some degree of subjectivity is inherent in this process.

We acknowledge that our use of fixed and separable ethical clusters represents a design choice that prioritizes clarity and tractability over capturing the full richness of moral reasoning. In practice, human decision-making rarely follows a single moral theory rigidly. Instead, individuals often blend elements from different ethical traditions depending on the context, and this hybrid reasoning may vary substantially across cultural or societal settings. For instance, care-oriented considerations may be weighted more heavily in collectivist societies, while deontological rules might dominate in legalistic contexts. In this paper we intentionally adopt separable clusters as a first step, as this provides a transparent mapping from well-established philosophical frameworks to computational representations, and makes belief aggregation methods easier to analyze. However, AMULED’s modular structure is designed to be extendable: future work could allow agents to dynamically adjust cluster weights, incorporate hybrid reasoning patterns, or learn culturally specific moral priors from human feedback. This flexibility would enable the framework to better mirror the nuanced and context-dependent way that people actually reason about ethics, while retaining the interpretability benefits of our current design.

Our framework is based on key assumptions: we believe that the five clusters effectively encompass the major streams of ethical thought pertinent to AI decision-making and that the selected theories within each cluster are sufficiently representative of their respective ethical approaches. We also assume that the identified key concepts and decision factors can be meaningfully translated into computational models, providing a solid foundation for future research. AMULED demonstrates strong generalization across diverse tasks like the static "Finding Milk" (grid-based planning with immediate ethics) and dynamic "Driving and Rescuing" (continuous motion, probabilistic elements, high-stakes decisions), highlighting its adaptability to varying environmental complexities and moral conflicts. The framework's task-agnostic ethical layer, integrating RL-LLM for scalable feedback, few-shot prompting for precise outputs, and multi-moral belief aggregation reduces dependence on hand-crafted rewards and balances primary goals with ethical sub-goals without domain-specific adjustments. Evidence from results shows consistent ethical alignment and superior outcomes over baselines in both tasks, suggesting robustness to moral uncertainty and potential extension to other domains (e.g., healthcare triage, resource allocation) via simple prompt and environment tweaks, without core architecture redesign.

Recognizing that different countries and cultures can value certain moral beliefs over another, AMULED was designed to have modular selection of ethical frameworks, rather than imposing a single, moral philosophy to drive decisions.
While we highlight that this modular, multi-moral approach avoids the pitfalls of sticking to a fixed moral framework, we do recognize that the representation of belief values generated from LLMs may be prone to external biases and subjectivity.
This complicates the design of a universally accepted modular decision-making framework. 
Moreover, the dynamic nature of ethics poses a challenge, as ethical norms can evolve over time and vary across cultures, potentially rendering static models ineffective. Accountability is another critical issue \cite{helbing2021summary}; ambiguity arises regarding who is responsible for the outcomes generated by AI, whether it be the developers, the organization deploying the AI, or the model itself. To build trust and accountability, enhancing the transparency of the AMULED model's decision-making processes is essential, potentially through techniques that provide explanations, visualizations, or justifications for its ethical reasoning. This approach would improve user confidence and ensure that the AI's ethical framework aligns with societal values and norms \cite{helbing2024converging}.


The integration of Belief Probability Assignment (BPA) into reinforcement learning offers a transformative approach to ethical decision-making in AI systems. By aggregating beliefs from multiple feedback sources, such as ethical LLMs and human evaluators, AI can effectively assess and balance competing moral frameworks. This capability is particularly valuable in contexts marked by normative uncertainty, where ethical models like utilitarianism and deontology may provide conflicting guidance. Through this aggregation, the AMULED system evaluates decisions with a nuanced understanding of diversity and confidence across ethical theories, enabling iterative refinement of strategies that align with widely accepted moral standards. Our framework explicitly accommodates multiple ethical perspectives, leveraging techniques such as Dempster–Shafer Theory (DST) to reconcile conflicts and generate coherent probabilistic outputs under uncertainty. While DST provides a mathematically principled way to combine beliefs, some subtleties of ethical tension may remain less transparent, highlighting opportunities for improving interpretability and accountability. Future work could focus on providing explainable justifications, tracking the influence of individual moral perspectives, and integrating contextual or culturally informed norms, thereby enhancing trust and practical applicability in dynamic, real-world environments, including multi-agent systems and human-AI collaborations.

While AMULED demonstrates strong performance in simulated environments, we see its greatest value as a methodological foundation for integrating pluralistic moral frameworks into RL. In this work, our focus was on establishing the technical feasibility and proof-of-concept behavior of the framework. Assessing how well AMULED aligns with human ethical reasoning is a natural next step rather than the primary goal of this paper. Looking ahead, we envision two complementary validation paths. One is user studies where participants evaluate AMULED-generated decisions in scenarios such as \texttt{FindMilk} and \texttt{Driving and Rescuing}, providing ratings of agreement and moral acceptability. Another is expert annotation, where ethicists or domain specialists assess a subset of AMULED’s decisions for consistency with philosophical frameworks and the clarity of their justification. By outlining these protocols, we position AMULED as both a step toward and a tool for empirical investigation of moral alignment, ensuring that future work can build directly on the methodological advances we present here.

\section{Conclusion and Future Works}
Our work advances the integration of ethical reasoning into AI, particularly in reinforcement learning frameworks, by leveraging ethically aligned LLM agents. While effective for low-stakes applications like personal or kitchen assistant robots, our approach is best viewed as complementary to human oversight, especially in high-stakes scenarios requiring rigorous testing. The framework excels in bootstrapping learning algorithms where human feedback is scarce, with the potential to replace LLM feedback with evaluations from multiple human evaluators, safeguarding against malicious manipulation.

Applications include smart home assistants, where AMULED promotes ethical and sustainable choices, and social media platforms, where it fosters constructive, inclusive interactions. However, categorizing ethical theories into clusters may oversimplify moral complexities, and operationalizing these theories involves abstracting philosophical principles. Our framework assumes that the selected ethical clusters and decision factors are representative and computationally translatable, though cultural and contextual variations pose challenges.

Future research should focus on hybrid ethical models, real-time adaptation, and interpretability to enhance trust and relevance. Addressing biases, exploring multi-agent scenarios, and establishing evaluation benchmarks are critical for equitable and robust outcomes. By integrating Belief Probability Assignment (BPA) into reinforcement learning, AMULED enables nuanced ethical decision-making, balancing competing moral frameworks in dynamic environments. Leveraging the Dempster-Shafer Theory for robust belief aggregation and the Jensen-Shannon Belief Divergence for precise information fusion, AMULED effectively synthesizes diverse ethical perspectives while quantifying uncertainty and resolving conflicts. These advanced techniques ensure that the system can adaptively weigh and harmonize competing moral values, such as utilitarianism and deontology, with a mathematically grounded approach. This paves the way for responsible, adaptable AI systems that are not only aligned with societal values but also capable of navigating the complexities of real-world ethical dilemmas with confidence and transparency.

\backmatter

\bibliography{refs}


\begin{thebibliography}{70}
\ifx \bisbn   \undefined \def \bisbn  #1{ISBN #1}\fi
\ifx \binits  \undefined \def \binits#1{#1}\fi
\ifx \bauthor  \undefined \def \bauthor#1{#1}\fi
\ifx \batitle  \undefined \def \batitle#1{#1}\fi
\ifx \bjtitle  \undefined \def \bjtitle#1{#1}\fi
\ifx \bvolume  \undefined \def \bvolume#1{\textbf{#1}}\fi
\ifx \byear  \undefined \def \byear#1{#1}\fi
\ifx \bissue  \undefined \def \bissue#1{#1}\fi
\ifx \bfpage  \undefined \def \bfpage#1{#1}\fi
\ifx \blpage  \undefined \def \blpage #1{#1}\fi
\ifx \burl  \undefined \def \burl#1{\textsf{#1}}\fi
\ifx \doiurl  \undefined \def \doiurl#1{\url{https://doi.org/#1}}\fi
\ifx \betal  \undefined \def \betal{\textit{et al.}}\fi
\ifx \binstitute  \undefined \def \binstitute#1{#1}\fi
\ifx \binstitutionaled  \undefined \def \binstitutionaled#1{#1}\fi
\ifx \bctitle  \undefined \def \bctitle#1{#1}\fi
\ifx \beditor  \undefined \def \beditor#1{#1}\fi
\ifx \bpublisher  \undefined \def \bpublisher#1{#1}\fi
\ifx \bbtitle  \undefined \def \bbtitle#1{#1}\fi
\ifx \bedition  \undefined \def \bedition#1{#1}\fi
\ifx \bseriesno  \undefined \def \bseriesno#1{#1}\fi
\ifx \blocation  \undefined \def \blocation#1{#1}\fi
\ifx \bsertitle  \undefined \def \bsertitle#1{#1}\fi
\ifx \bsnm \undefined \def \bsnm#1{#1}\fi
\ifx \bsuffix \undefined \def \bsuffix#1{#1}\fi
\ifx \bparticle \undefined \def \bparticle#1{#1}\fi
\ifx \barticle \undefined \def \barticle#1{#1}\fi
\bibcommenthead
\ifx \bconfdate \undefined \def \bconfdate #1{#1}\fi
\ifx \botherref \undefined \def \botherref #1{#1}\fi
\ifx \url \undefined \def \url#1{\textsf{#1}}\fi
\ifx \bchapter \undefined \def \bchapter#1{#1}\fi
\ifx \bbook \undefined \def \bbook#1{#1}\fi
\ifx \bcomment \undefined \def \bcomment#1{#1}\fi
\ifx \oauthor \undefined \def \oauthor#1{#1}\fi
\ifx \citeauthoryear \undefined \def \citeauthoryear#1{#1}\fi
\ifx \endbibitem  \undefined \def \endbibitem {}\fi
\ifx \bconflocation  \undefined \def \bconflocation#1{#1}\fi
\ifx \arxivurl  \undefined \def \arxivurl#1{\textsf{#1}}\fi
\csname PreBibitemsHook\endcsname

\bibitem[\protect\citeauthoryear{Chen}{2024}]{chen2024lost}
\begin{barticle}
\bauthor{\bsnm{Chen}, \binits{S.}}:
\batitle{The lost data: how {AI} systems censor {LGBTQ+} content in the name of
  safety}.
\bjtitle{Nature computational science}
\bvolume{4}(\bissue{9}),
\bfpage{629}--\blpage{632}
(\byear{2024})
\doiurl{10.1038/s43588-024-00695-4}
\end{barticle}
\endbibitem

\bibitem[\protect\citeauthoryear{Allen et~al.}{2000}]{allen2000prolegomena}
\begin{barticle}
\bauthor{\bsnm{Allen}, \binits{C.}},
\bauthor{\bsnm{Varner}, \binits{G.}},
\bauthor{\bsnm{Zinser}, \binits{J.}}:
\batitle{Prolegomena to any future artificial moral agent}.
\bjtitle{Journal of Experimental \& Theoretical Artificial Intelligence}
\bvolume{12}(\bissue{3}),
\bfpage{251}--\blpage{261}
(\byear{2000})
\end{barticle}
\endbibitem

\bibitem[\protect\citeauthoryear{Wallach and Allen}{2008}]{wallach2008moral}
\begin{bbook}
\bauthor{\bsnm{Wallach}, \binits{W.}},
\bauthor{\bsnm{Allen}, \binits{C.}}:
\bbtitle{Moral Machines: Teaching Robots Right from Wrong},
(\byear{2008})
\end{bbook}
\endbibitem

\bibitem[\protect\citeauthoryear{Bringsjord
  et~al.}{2006}]{bringsjord2006toward}
\begin{barticle}
\bauthor{\bsnm{Bringsjord}, \binits{S.}},
\bauthor{\bsnm{Arkoudas}, \binits{K.}},
\bauthor{\bsnm{Bello}, \binits{P.}}:
\batitle{Toward a general logicist methodology for engineering ethically
  correct robots}.
\bjtitle{IEEE Intelligent Systems}
\bvolume{21}(\bissue{4}),
\bfpage{38}--\blpage{44}
(\byear{2006})
\doiurl{10.1109/mis.2006.82}
\end{barticle}
\endbibitem

\bibitem[\protect\citeauthoryear{Russell et~al.}{2015}]{russell2015research}
\begin{barticle}
\bauthor{\bsnm{Russell}, \binits{S.}},
\bauthor{\bsnm{Dewey}, \binits{D.}},
\bauthor{\bsnm{Tegmark}, \binits{M.}}:
\batitle{Research priorities for robust and beneficial artificial
  intelligence}.
\bjtitle{AI magazine}
\bvolume{36}(\bissue{4}),
\bfpage{105}--\blpage{114}
(\byear{2015})
\end{barticle}
\endbibitem

\bibitem[\protect\citeauthoryear{Ecoffet and
  Lehman}{2020}]{ecoffet2020moraluncertainty}
\begin{botherref}
\oauthor{\bsnm{Ecoffet}, \binits{A.}},
\oauthor{\bsnm{Lehman}, \binits{J.}}:
Reinforcement Learning Under Moral Uncertainty.
arXiv
(2020).
\doiurl{10.48550/arxiv.2006.04734}
\end{botherref}
\endbibitem

\bibitem[\protect\citeauthoryear{MacAskill et~al.}{2020}]{macaskill2020moral}
\begin{bbook}
\bauthor{\bsnm{MacAskill}, \binits{M.}},
\bauthor{\bsnm{Bykvist}, \binits{K.}},
\bauthor{\bsnm{Ord}, \binits{T.}}:
\bbtitle{Moral Uncertainty}.
\bpublisher{Oxford University Press}, \blocation{???}
(\byear{2020}).
\doiurl{10.1093/oso/9780198722274.001.0001}
\end{bbook}
\endbibitem

\bibitem[\protect\citeauthoryear{L{\"u}tge}{2019}]{lutge2019ethics}
\begin{bbook}
\bauthor{\bsnm{L{\"u}tge}, \binits{C.}}:
\bbtitle{The Ethics of Competition: How a Competitive Society Is Good for All}.
\bpublisher{Edward Elgar Publishing}, \blocation{???}
(\byear{2019})
\end{bbook}
\endbibitem

\bibitem[\protect\citeauthoryear{Lindner and
  Bentzen}{2018}]{lindner2018formalization}
\begin{botherref}
\oauthor{\bsnm{Lindner}, \binits{F.}},
\oauthor{\bsnm{Bentzen}, \binits{M.M.}}:
A formalization of kant's second formulation of the categorical imperative.
arXiv preprint arXiv:1801.03160
(2018)
\end{botherref}
\endbibitem

\bibitem[\protect\citeauthoryear{Abel et~al.}{2016}]{abel2016reinforcement}
\begin{bchapter}
\bauthor{\bsnm{Abel}, \binits{D.}},
\bauthor{\bsnm{MacGlashan}, \binits{J.}},
\bauthor{\bsnm{Littman}, \binits{M.L.}}:
\bctitle{Reinforcement learning as a framework for ethical decision making}.
In: \bbtitle{Workshops at the Thirtieth {AAAI} Conference on Artificial
  Intelligence}
(\byear{2016})
\end{bchapter}
\endbibitem

\bibitem[\protect\citeauthoryear{Hadfield-Menell
  et~al.}{2017}]{hadfield2017off}
\begin{bchapter}
\bauthor{\bsnm{Hadfield-Menell}, \binits{D.}},
\bauthor{\bsnm{Dragan}, \binits{A.}},
\bauthor{\bsnm{Abbeel}, \binits{P.}},
\bauthor{\bsnm{Russell}, \binits{S.}}:
\bctitle{The off-switch game}.
In: \bbtitle{Workshops at the Thirty-First {AAAI} Conference on Artificial
  Intelligence},
pp. \bfpage{220}--\blpage{227}.
\bpublisher{International Joint Conferences on Artificial Intelligence
  Organization}, \blocation{???}
(\byear{2017}).
\doiurl{10.24963/ijcai.2017/32}
\end{bchapter}
\endbibitem

\bibitem[\protect\citeauthoryear{Ng et~al.}{2000}]{ng2000algorithms}
\begin{bchapter}
\bauthor{\bsnm{Ng}, \binits{A.Y.}},
\bauthor{\bsnm{Russell}, \binits{S.}}, \betal:
\bctitle{Algorithms for inverse reinforcement learning.}
In: \bbtitle{ICML},
vol. \bseriesno{1},
p. \bfpage{2}
(\byear{2000})
\end{bchapter}
\endbibitem

\bibitem[\protect\citeauthoryear{Ziebart et~al.}{2008}]{ziebart2008maximum}
\begin{bchapter}
\bauthor{\bsnm{Ziebart}, \binits{B.D.}},
\bauthor{\bsnm{Maas}, \binits{A.L.}},
\bauthor{\bsnm{Bagnell}, \binits{J.A.}},
\bauthor{\bsnm{Dey}, \binits{A.K.}}, \betal:
\bctitle{Maximum entropy inverse reinforcement learning.}
In: \bbtitle{{AAAI}},
vol. \bseriesno{8},
pp. \bfpage{1433}--\blpage{1438}
(\byear{2008}).
\bcomment{Chicago, IL, USA}
\end{bchapter}
\endbibitem

\bibitem[\protect\citeauthoryear{Malle}{2016}]{malle2016integrating}
\begin{barticle}
\bauthor{\bsnm{Malle}, \binits{B.F.}}:
\batitle{Integrating robot ethics and machine morality: the study and design of
  moral competence in robots}.
\bjtitle{Ethics and Information Technology}
\bvolume{18},
\bfpage{243}--\blpage{256}
(\byear{2016})
\doiurl{10.1007/s10676-015-9367-8}
\end{barticle}
\endbibitem

\bibitem[\protect\citeauthoryear{Dennis et~al.}{2016}]{dennis2016formal}
\begin{barticle}
\bauthor{\bsnm{Dennis}, \binits{L.}},
\bauthor{\bsnm{Fisher}, \binits{M.}},
\bauthor{\bsnm{Slavkovik}, \binits{M.}},
\bauthor{\bsnm{Webster}, \binits{M.}}:
\batitle{Formal verification of ethical choices in autonomous systems}.
\bjtitle{Robotics and Autonomous Systems}
\bvolume{77},
\bfpage{1}--\blpage{14}
(\byear{2016})
\doiurl{10.1016/j.robot.2015.11.012}
\end{barticle}
\endbibitem

\bibitem[\protect\citeauthoryear{Santoni~de Sio and Van~den
  Hoven}{2018}]{santoni2018meaningful}
\begin{barticle}
\bauthor{\bsnm{Sio}, \binits{F.}},
\bauthor{\bsnm{Hoven}, \binits{J.}}:
\batitle{Meaningful human control over autonomous systems: A philosophical
  account}.
\bjtitle{Frontiers in Robotics and AI}
\bvolume{5},
\bfpage{15}
(\byear{2018})
\doiurl{10.3389/frobt.2018.00015}
\end{barticle}
\endbibitem

\bibitem[\protect\citeauthoryear{Cummings}{2014}]{cummings2014man}
\begin{barticle}
\bauthor{\bsnm{Cummings}, \binits{M.M.}}:
\batitle{Man versus machine or man+ machine?}
\bjtitle{IEEE Intelligent Systems}
\bvolume{29}(\bissue{5}),
\bfpage{62}--\blpage{69}
(\byear{2014})
\doiurl{10.1109/mis.2014.87}
\end{barticle}
\endbibitem

\bibitem[\protect\citeauthoryear{Suri}{2024}]{suri2024defining}
\begin{barticle}
\bauthor{\bsnm{Suri}, \binits{S.}}:
\batitle{Defining our future with generative {AI}}.
\bjtitle{Nature Computational Science}
\bvolume{4}(\bissue{9}),
\bfpage{641}--\blpage{643}
(\byear{2024})
\doiurl{10.1038/s43588-024-00694-5}
\end{barticle}
\endbibitem

\bibitem[\protect\citeauthoryear{Van~Noorden and Perkel}{2023}]{van2023ai}
\begin{barticle}
\bauthor{\bsnm{Van~Noorden}, \binits{R.}},
\bauthor{\bsnm{Perkel}, \binits{J.M.}}:
\batitle{Ai and science: what 1,600 researchers think}.
\bjtitle{Nature}
\bvolume{621}(\bissue{7980}),
\bfpage{672}--\blpage{675}
(\byear{2023})
\doiurl{10.1038/d41586-023-02980-0}
\end{barticle}
\endbibitem

\bibitem[\protect\citeauthoryear{Park et~al.}{2023}]{park_generative_2023}
\begin{botherref}
\oauthor{\bsnm{Park}, \binits{J.S.}},
\oauthor{\bsnm{O'Brien}, \binits{J.C.}},
\oauthor{\bsnm{Cai}, \binits{C.J.}},
\oauthor{\bsnm{Morris}, \binits{M.R.}},
\oauthor{\bsnm{Liang}, \binits{P.}},
\oauthor{\bsnm{Bernstein}, \binits{M.S.}}:
{Generative Agents: Interactive Simulacra of Human Behavior}.
arXiv
(2023).
\doiurl{10.48550/arxiv.2304.03442}
\end{botherref}
\endbibitem

\bibitem[\protect\citeauthoryear{Yang et~al.}{2024}]{yang2024llmvoting}
\begin{barticle}
\bauthor{\bsnm{Yang}, \binits{J.C.}},
\bauthor{\bsnm{Dailisan}, \binits{D.}},
\bauthor{\bsnm{Korecki}, \binits{M.}},
\bauthor{\bsnm{Hausladen}, \binits{C.I.}},
\bauthor{\bsnm{Helbing}, \binits{D.}}:
\batitle{{LLM} voting: Human choices and {AI} collective decision making}.
\bjtitle{Proceedings of the AAAI/ACM Conference on AI, Ethics, and Society}
\bvolume{7},
\bfpage{1696}--\blpage{1708}
(\byear{2024})
\doiurl{10.1609/aies.v7i1.31758}
\end{barticle}
\endbibitem

\bibitem[\protect\citeauthoryear{Gudi{\~n}o-Rosero
  et~al.}{2024}]{gudino2024large}
\begin{botherref}
\oauthor{\bsnm{Gudi{\~n}o-Rosero}, \binits{J.}},
\oauthor{\bsnm{Grandi}, \binits{U.}},
\oauthor{\bsnm{Hidalgo}, \binits{C.A.}}:
Large language models ({LLM}s) as agents for augmented democracy.
Philosophical Transactions of the Royal Society A: Mathematical, Physical and
  Engineering Sciences
\textbf{382}(2285)
(2024)
\doiurl{10.1098/rsta.2024.0100}
\end{botherref}
\endbibitem

\bibitem[\protect\citeauthoryear{Wang et~al.}{2024}]{wang2024survey}
\begin{botherref}
\oauthor{\bsnm{Wang}, \binits{L.}},
\oauthor{\bsnm{Ma}, \binits{C.}},
\oauthor{\bsnm{Feng}, \binits{X.}},
\oauthor{\bsnm{Zhang}, \binits{Z.}},
\oauthor{\bsnm{Yang}, \binits{H.}},
\oauthor{\bsnm{Zhang}, \binits{J.}},
\oauthor{\bsnm{Chen}, \binits{Z.}},
\oauthor{\bsnm{Tang}, \binits{J.}},
\oauthor{\bsnm{Chen}, \binits{X.}},
\oauthor{\bsnm{Lin}, \binits{Y.}},
\oauthor{\bsnm{Zhao}, \binits{W.X.}},
\oauthor{\bsnm{Wei}, \binits{Z.}},
\oauthor{\bsnm{Wen}, \binits{J.}}:
A survey on large language model based autonomous agents.
Frontiers of Computer Science
\textbf{18}(6)
(2024)
\doiurl{10.1007/s11704-024-40231-1}
\end{botherref}
\endbibitem

\bibitem[\protect\citeauthoryear{Scherrer
  et~al.}{2023}]{scherrer2023moralbeliefllm}
\begin{bchapter}
\bauthor{\bsnm{Scherrer}, \binits{N.}},
\bauthor{\bsnm{Shi}, \binits{C.}},
\bauthor{\bsnm{Feder}, \binits{A.}},
\bauthor{\bsnm{Blei}, \binits{D.}}:
\bctitle{Evaluating the moral beliefs encoded in {LLM}s}.
In: \beditor{\bsnm{Oh}, \binits{A.}},
\beditor{\bsnm{Naumann}, \binits{T.}},
\beditor{\bsnm{Globerson}, \binits{A.}},
\beditor{\bsnm{Saenko}, \binits{K.}},
\beditor{\bsnm{Hardt}, \binits{M.}},
\beditor{\bsnm{Levine}, \binits{S.}} (eds.)
\bbtitle{Advances in Neural Information Processing Systems},
vol. \bseriesno{36},
pp. \bfpage{51778}--\blpage{51809}
(\byear{2023})
\end{bchapter}
\endbibitem

\bibitem[\protect\citeauthoryear{Garcia et~al.}{2024}]{garcia2024moralturing}
\begin{botherref}
\oauthor{\bsnm{Garcia}, \binits{B.}},
\oauthor{\bsnm{Qian}, \binits{C.}},
\oauthor{\bsnm{Palminteri}, \binits{S.}}:
The Moral Turing Test: Evaluating Human-{LLM} Alignment in Moral
  Decision-Making.
arXiv
(2024).
\doiurl{10.48550/arxiv.2410.07304}
\end{botherref}
\endbibitem

\bibitem[\protect\citeauthoryear{Sen}{2017}]{sen2017collective}
\begin{bbook}
\bauthor{\bsnm{Sen}, \binits{A.}}:
\bbtitle{Collective Choice and Social Welfare: Expanded Edition}.
\bpublisher{Penguin UK}, \blocation{???}
(\byear{2017})
\end{bbook}
\endbibitem

\bibitem[\protect\citeauthoryear{Hagendorff et~al.}{2023}]{hagendorff2023human}
\begin{barticle}
\bauthor{\bsnm{Hagendorff}, \binits{T.}},
\bauthor{\bsnm{Fabi}, \binits{S.}},
\bauthor{\bsnm{Kosinski}, \binits{M.}}:
\batitle{Human-like intuitive behavior and reasoning biases emerged in large
  language models but disappeared in {ChatGPT}}.
\bjtitle{Nature Computational Science}
\bvolume{3}(\bissue{10}),
\bfpage{833}--\blpage{838}
(\byear{2023})
\doiurl{10.1038/s43588-023-00527-x}
\end{barticle}
\endbibitem

\bibitem[\protect\citeauthoryear{Gips}{2011}]{gips2011towards}
\begin{barticle}
\bauthor{\bsnm{Gips}, \binits{J.}}:
\batitle{Towards the ethical robot}.
\bjtitle{Machine ethics}
\bvolume{1},
\bfpage{244}--\blpage{253}
(\byear{2011})
\doiurl{10.1017/cbo9780511978036.019}
\end{barticle}
\endbibitem

\bibitem[\protect\citeauthoryear{Jiao et~al.}{2024}]{jiao2024navigating}
\begin{botherref}
\oauthor{\bsnm{Jiao}, \binits{J.}},
\oauthor{\bsnm{Afroogh}, \binits{S.}},
\oauthor{\bsnm{Xu}, \binits{Y.}},
\oauthor{\bsnm{Phillips}, \binits{C.}}:
Navigating {LLM} ethics: Advancements, challenges, and future directions.
arXiv preprint arXiv:2406.18841
(2024)
\end{botherref}
\endbibitem

\bibitem[\protect\citeauthoryear{Coeckelbergh}{2025}]{coeckelbergh2025llms}
\begin{barticle}
\bauthor{\bsnm{Coeckelbergh}, \binits{M.}}:
\batitle{{LLM}s, truth, and democracy: An overview of risks}.
\bjtitle{Science and Engineering Ethics}
\bvolume{31}(\bissue{1}),
\bfpage{1}--\blpage{13}
(\byear{2025})
\end{barticle}
\endbibitem

\bibitem[\protect\citeauthoryear{Xie et~al.}{2024}]{xie2024fire}
\begin{botherref}
\oauthor{\bsnm{Xie}, \binits{Z.}},
\oauthor{\bsnm{Xing}, \binits{R.}},
\oauthor{\bsnm{Wang}, \binits{Y.}},
\oauthor{\bsnm{Geng}, \binits{J.}},
\oauthor{\bsnm{Iqbal}, \binits{H.}},
\oauthor{\bsnm{Sahnan}, \binits{D.}},
\oauthor{\bsnm{Gurevych}, \binits{I.}},
\oauthor{\bsnm{Nakov}, \binits{P.}}:
Fire: Fact-checking with iterative retrieval and verification.
arXiv preprint arXiv:2411.00784
(2024)
\end{botherref}
\endbibitem

\bibitem[\protect\citeauthoryear{Achiam et~al.}{2017}]{achiam2017constrained}
\begin{bchapter}
\bauthor{\bsnm{Achiam}, \binits{J.}},
\bauthor{\bsnm{Held}, \binits{D.}},
\bauthor{\bsnm{Tamar}, \binits{A.}},
\bauthor{\bsnm{Abbeel}, \binits{P.}}:
\bctitle{Constrained policy optimization}.
In: \bbtitle{Proceedings of the 34th International Conference on Machine
  Learning},
pp. \bfpage{22}--\blpage{31}
(\byear{2017}).
\bcomment{PMLR}
\end{bchapter}
\endbibitem

\bibitem[\protect\citeauthoryear{Tamar et~al.}{2015}]{tamar2015policy}
\begin{bchapter}
\bauthor{\bsnm{Tamar}, \binits{A.}},
\bauthor{\bsnm{Chow}, \binits{Y.}},
\bauthor{\bsnm{Ghavamzadeh}, \binits{M.}},
\bauthor{\bsnm{Mannor}, \binits{S.}}:
\bctitle{Policy gradients with variance related risk criteria}.
In: \bbtitle{Advances in Neural Information Processing Systems},
pp. \bfpage{167}--\blpage{175}
(\byear{2015})
\end{bchapter}
\endbibitem

\bibitem[\protect\citeauthoryear{Chow et~al.}{2018}]{chow2018lyapunov}
\begin{bchapter}
\bauthor{\bsnm{Chow}, \binits{Y.}},
\bauthor{\bsnm{Nachum}, \binits{O.}},
\bauthor{\bsnm{Du{\'e}nez-Guzm{\'a}n}, \binits{E.}},
\bauthor{\bsnm{Ghavamzadeh}, \binits{M.}}:
\bctitle{Lyapunov-based safe policy optimization for continuous control}.
In: \bbtitle{Proceedings of the 35th International Conference on Machine
  Learning},
pp. \bfpage{809}--\blpage{818}
(\byear{2018}).
\bcomment{PMLR}
\end{bchapter}
\endbibitem

\bibitem[\protect\citeauthoryear{Alshiekh et~al.}{2018}]{alshiekh2018safe}
\begin{bchapter}
\bauthor{\bsnm{Alshiekh}, \binits{M.}},
\bauthor{\bsnm{Bloem}, \binits{R.}},
\bauthor{\bsnm{Ehlers}, \binits{R.}},
\bauthor{\bsnm{Koch}, \binits{M.}},
\bauthor{\bsnm{K{\"o}nighofer}, \binits{B.}},
\bauthor{\bsnm{Niekum}, \binits{S.}},
\bauthor{\bsnm{Topcu}, \binits{U.}}:
\bctitle{Safe reinforcement learning via shielding}.
In: \bbtitle{Proceedings of the AAAI Conference on Artificial Intelligence},
vol. \bseriesno{32}
(\byear{2018})
\end{bchapter}
\endbibitem

\bibitem[\protect\citeauthoryear{Berkenkamp et~al.}{2017}]{berkenkamp2017safe}
\begin{bchapter}
\bauthor{\bsnm{Berkenkamp}, \binits{F.}},
\bauthor{\bsnm{Turchetta}, \binits{M.}},
\bauthor{\bsnm{Schoellig}, \binits{A.P.}},
\bauthor{\bsnm{Krause}, \binits{A.}}:
\bctitle{Safe model-based reinforcement learning with stability guarantees}.
In: \bbtitle{Advances in Neural Information Processing Systems},
pp. \bfpage{908}--\blpage{918}
(\byear{2017})
\end{bchapter}
\endbibitem

\bibitem[\protect\citeauthoryear{Morimoto and Doya}{2005}]{morimoto2005robust}
\begin{barticle}
\bauthor{\bsnm{Morimoto}, \binits{J.}},
\bauthor{\bsnm{Doya}, \binits{K.}}:
\batitle{Robust reinforcement learning}.
\bjtitle{Neural Computation}
\bvolume{17}(\bissue{2}),
\bfpage{335}--\blpage{359}
(\byear{2005})
\end{barticle}
\endbibitem

\bibitem[\protect\citeauthoryear{Garc{\'\i}a and
  Fern{\'a}ndez}{2015}]{garcia2015comprehensive}
\begin{barticle}
\bauthor{\bsnm{Garc{\'\i}a}, \binits{J.}},
\bauthor{\bsnm{Fern{\'a}ndez}, \binits{F.}}:
\batitle{A comprehensive survey on safe reinforcement learning}.
\bjtitle{Journal of Machine Learning Research}
\bvolume{16}(\bissue{1}),
\bfpage{1437}--\blpage{1480}
(\byear{2015})
\end{barticle}
\endbibitem

\bibitem[\protect\citeauthoryear{Gu et~al.}{2022}]{gu2022review}
\begin{botherref}
\oauthor{\bsnm{Gu}, \binits{S.}},
\oauthor{\bsnm{Yang}, \binits{L.}},
\oauthor{\bsnm{Du}, \binits{Y.}},
\oauthor{\bsnm{Chen}, \binits{G.}},
\oauthor{\bsnm{Walter}, \binits{F.}},
\oauthor{\bsnm{Wang}, \binits{J.}},
\oauthor{\bsnm{Knoll}, \binits{A.}}:
A review of safe reinforcement learning: Methods, theory and applications.
arXiv preprint arXiv:2205.10330
(2022)
\end{botherref}
\endbibitem

\bibitem[\protect\citeauthoryear{Duan et~al.}{2019}]{duan2019artificial}
\begin{barticle}
\bauthor{\bsnm{Duan}, \binits{Y.}},
\bauthor{\bsnm{Edwards}, \binits{J.S.}},
\bauthor{\bsnm{Dwivedi}, \binits{Y.K.}}:
\batitle{Artificial intelligence for decision making in the era of big
  data--evolution, challenges and research agenda}.
\bjtitle{International journal of information management}
\bvolume{48},
\bfpage{63}--\blpage{71}
(\byear{2019})
\doiurl{10.1016/j.ijinfomgt.2019.01.021}
\end{barticle}
\endbibitem

\bibitem[\protect\citeauthoryear{Mahajan}{2024}]{mahajan2024executioner}
\begin{botherref}
\oauthor{\bsnm{Mahajan}, \binits{S.}}:
The executioner paradox: understanding self-referential dilemma in
  computational systems.
AI \& SOCIETY,
1--8
(2024)
\doiurl{10.1007/s00146-024-01968-2}
\end{botherref}
\endbibitem

\bibitem[\protect\citeauthoryear{Tolmeijer
  et~al.}{2020}]{tolmeijer2020implementations}
\begin{barticle}
\bauthor{\bsnm{Tolmeijer}, \binits{S.}},
\bauthor{\bsnm{Kneer}, \binits{M.}},
\bauthor{\bsnm{Sarasua}, \binits{C.}},
\bauthor{\bsnm{Christen}, \binits{M.}},
\bauthor{\bsnm{Bernstein}, \binits{A.}}:
\batitle{Implementations in machine ethics: A survey}.
\bjtitle{ACM Computing Surveys (CSUR)}
\bvolume{53}(\bissue{6}),
\bfpage{1}--\blpage{38}
(\byear{2020})
\doiurl{10.1145/3419633}
\end{barticle}
\endbibitem

\bibitem[\protect\citeauthoryear{Schramowski
  et~al.}{2022}]{schramowski2022large}
\begin{barticle}
\bauthor{\bsnm{Schramowski}, \binits{P.}},
\bauthor{\bsnm{Turan}, \binits{C.}},
\bauthor{\bsnm{Andersen}, \binits{N.}},
\bauthor{\bsnm{Rothkopf}, \binits{C.A.}},
\bauthor{\bsnm{Kersting}, \binits{K.}}:
\batitle{Large pre-trained language models contain human-like biases of what is
  right and wrong to do}.
\bjtitle{Nature Machine Intelligence}
\bvolume{4}(\bissue{3}),
\bfpage{258}--\blpage{268}
(\byear{2022})
\end{barticle}
\endbibitem

\bibitem[\protect\citeauthoryear{Anderson and
  Anderson}{2011}]{anderson2011machine}
\begin{bbook}
\bauthor{\bsnm{Anderson}, \binits{M.}},
\bauthor{\bsnm{Anderson}, \binits{S.L.}}:
\bbtitle{Machine Ethics}.
\bpublisher{Cambridge University Press}, \blocation{???}
(\byear{2011}).
\doiurl{10.1017/cbo9780511978036}
\end{bbook}
\endbibitem

\bibitem[\protect\citeauthoryear{Powers}{2006}]{powers2006prospects}
\begin{barticle}
\bauthor{\bsnm{Powers}, \binits{T.M.}}:
\batitle{Prospects for a kantian machine}.
\bjtitle{IEEE Intelligent Systems}
\bvolume{21}(\bissue{4}),
\bfpage{46}--\blpage{51}
(\byear{2006})
\end{barticle}
\endbibitem

\bibitem[\protect\citeauthoryear{Alexander and
  Moore}{2024}]{alexander2007deontological}
\begin{bchapter}
\bauthor{\bsnm{Alexander}, \binits{L.}},
\bauthor{\bsnm{Moore}, \binits{M.}}:
\bctitle{{Deontological Ethics}}.
In: \beditor{\bsnm{Zalta}, \binits{E.N.}},
\beditor{\bsnm{Nodelman}, \binits{U.}} (eds.)
\bbtitle{The {Stanford} Encyclopedia of Philosophy},
\bedition{{W}inter 2024} edn.
\bpublisher{Metaphysics Research Lab, Stanford University}, \blocation{???}
(\byear{2024})
\end{bchapter}
\endbibitem

\bibitem[\protect\citeauthoryear{Gilligan}{2014}]{gilligan2014moral}
\begin{barticle}
\bauthor{\bsnm{Gilligan}, \binits{C.}}:
\batitle{Moral injury and the ethic of care: Reframing the conversation about
  differences.}
\bjtitle{Journal of social philosophy}
\bvolume{45}(\bissue{1}),
\bfpage{89}--\blpage{106}
(\byear{2014})
\end{barticle}
\endbibitem

\bibitem[\protect\citeauthoryear{Pogge}{2007}]{pogge2007john}
\begin{botherref}
\oauthor{\bsnm{Pogge}, \binits{T.}}:
John Rawls: His Life and Theory of Justice.
Oxford University Press
(2007).
\doiurl{10.5860/choice.45-1128}
\end{botherref}
\endbibitem

\bibitem[\protect\citeauthoryear{Dignum}{2018}]{dignum2018ethics}
\begin{barticle}
\bauthor{\bsnm{Dignum}, \binits{V.}}:
\batitle{Ethics in artificial intelligence: introduction to the special issue}.
\bjtitle{Ethics and Information Technology}
\bvolume{20}(\bissue{1}),
\bfpage{1}--\blpage{3}
(\byear{2018})
\doiurl{10.1007/s10676-018-9450-z}
\end{barticle}
\endbibitem

\bibitem[\protect\citeauthoryear{Conitzer et~al.}{2017}]{conitzer2017moral}
\begin{bchapter}
\bauthor{\bsnm{Conitzer}, \binits{V.}},
\bauthor{\bsnm{Sinnott-Armstrong}, \binits{W.}},
\bauthor{\bsnm{Borg}, \binits{J.S.}},
\bauthor{\bsnm{Deng}, \binits{Y.}},
\bauthor{\bsnm{Kramer}, \binits{M.}}:
\bctitle{Moral decision making frameworks for artificial intelligence}.
In: \bbtitle{Proceedings of the {AAAI} Conference on Artificial Intelligence},
vol. \bseriesno{31}.
\bpublisher{Association for the Advancement of Artificial Intelligence (AAAI)},
  \blocation{???}
(\byear{2017}).
\doiurl{10.1609/aaai.v31i1.11140}
\end{bchapter}
\endbibitem

\bibitem[\protect\citeauthoryear{Bench-Capon}{2020}]{bench2020ethical}
\begin{barticle}
\bauthor{\bsnm{Bench-Capon}, \binits{T.J.}}:
\batitle{Ethical approaches and autonomous systems}.
\bjtitle{Artificial Intelligence}
\bvolume{281},
\bfpage{103239}
(\byear{2020})
\end{barticle}
\endbibitem

\bibitem[\protect\citeauthoryear{Chen and Deng}{2023}]{Chen2023}
\begin{barticle}
\bauthor{\bsnm{Chen}, \binits{X.}},
\bauthor{\bsnm{Deng}, \binits{Y.}}:
\batitle{A novel combination rule for conflict management in data fusion}.
\bjtitle{Soft Computing}
\bvolume{27}(\bissue{22}),
\bfpage{16483}--\blpage{16492}
(\byear{2023})
\doiurl{10.1007/s00500-023-09112-w}
\end{barticle}
\endbibitem

\bibitem[\protect\citeauthoryear{Zhao et~al.}{2022}]{Zhao2022}
\begin{barticle}
\bauthor{\bsnm{Zhao}, \binits{K.}},
\bauthor{\bsnm{Li}, \binits{L.}},
\bauthor{\bsnm{Chen}, \binits{Z.}},
\bauthor{\bsnm{Sun}, \binits{R.}},
\bauthor{\bsnm{Yuan}, \binits{G.}},
\bauthor{\bsnm{Li}, \binits{J.}}:
\batitle{A survey: Optimization and applications of evidence fusion algorithm
  based on {Dempster--Shafer} theory}.
\bjtitle{Applied Soft Computing}
\bvolume{124},
\bfpage{109075}
(\byear{2022})
\doiurl{10.1016/j.asoc.2022.109075}
\end{barticle}
\endbibitem

\bibitem[\protect\citeauthoryear{Xiao}{2019}]{xiao2019multi}
\begin{barticle}
\bauthor{\bsnm{Xiao}, \binits{F.}}:
\batitle{Multi-sensor data fusion based on the belief divergence measure of
  evidences and the belief entropy}.
\bjtitle{Information Fusion}
\bvolume{46},
\bfpage{23}--\blpage{32}
(\byear{2019})
\doiurl{10.1016/j.inffus.2018.04.003}
\end{barticle}
\endbibitem

\bibitem[\protect\citeauthoryear{Dempster}{2008}]{dempster2008upper}
\begin{bchapter}
\bauthor{\bsnm{Dempster}, \binits{A.P.}}:
\bctitle{Upper and lower probabilities induced by a multivalued mapping}.
In: \bbtitle{Classic Works of the Dempster-Shafer Theory of Belief Functions},
pp. \bfpage{57}--\blpage{72}.
\bpublisher{Springer}, \blocation{???}
(\byear{2008}).
\doiurl{10.1007/978-3-540-44792-4_3}
\end{bchapter}
\endbibitem

\bibitem[\protect\citeauthoryear{MacAskill}{2014}]{macaskill2014normative}
\begin{botherref}
\oauthor{\bsnm{MacAskill}, \binits{W.}}:
Normative uncertainty.
PhD thesis,
University of Oxford
(2014)
\end{botherref}
\endbibitem

\bibitem[\protect\citeauthoryear{Ugazio et~al.}{2021}]{10.1093/scan/nsab100}
\begin{barticle}
\bauthor{\bsnm{Ugazio}, \binits{G.}},
\bauthor{\bsnm{Grueschow}, \binits{M.}},
\bauthor{\bsnm{Polania}, \binits{R.}},
\bauthor{\bsnm{Lamm}, \binits{C.}},
\bauthor{\bsnm{Tobler}, \binits{P.}},
\bauthor{\bsnm{Ruff}, \binits{C.}}:
\batitle{{Neuro-computational foundations of moral preferences}}.
\bjtitle{Social Cognitive and Affective Neuroscience}
\bvolume{17}(\bissue{3}),
\bfpage{253}--\blpage{265}
(\byear{2021})
\doiurl{10.1093/scan/nsab100}
\end{barticle}
\endbibitem

\bibitem[\protect\citeauthoryear{Wei et~al.}{2022}]{wei2022chainofthought}
\begin{bchapter}
\bauthor{\bsnm{Wei}, \binits{J.}},
\bauthor{\bsnm{Wang}, \binits{X.}},
\bauthor{\bsnm{Schuurmans}, \binits{D.}},
\bauthor{\bsnm{Bosma}, \binits{M.}},
\bauthor{\bsnm{ichter}, \binits{b.}},
\bauthor{\bsnm{Xia}, \binits{F.}},
\bauthor{\bsnm{Chi}, \binits{E.}},
\bauthor{\bsnm{Le}, \binits{Q.V.}},
\bauthor{\bsnm{Zhou}, \binits{D.}}:
\bctitle{Chain-of-thought prompting elicits reasoning in large language
  models}.
In: \beditor{\bsnm{Koyejo}, \binits{S.}},
\beditor{\bsnm{Mohamed}, \binits{S.}},
\beditor{\bsnm{Agarwal}, \binits{A.}},
\beditor{\bsnm{Belgrave}, \binits{D.}},
\beditor{\bsnm{Cho}, \binits{K.}},
\beditor{\bsnm{Oh}, \binits{A.}} (eds.)
\bbtitle{Advances in Neural Information Processing Systems},
vol. \bseriesno{35},
pp. \bfpage{24824}--\blpage{24837}
(\byear{2022})
\end{bchapter}
\endbibitem

\bibitem[\protect\citeauthoryear{Bellman}{1957}]{bellman1957markovian}
\begin{barticle}
\bauthor{\bsnm{Bellman}, \binits{R.}}:
\batitle{A markovian decision process}.
\bjtitle{J. Math. Mech.}
\bvolume{6},
\bfpage{679}--\blpage{684}
(\byear{1957})
\end{barticle}
\endbibitem

\bibitem[\protect\citeauthoryear{Sutton and
  Barto}{2018}]{sutton2018reinforcement}
\begin{bbook}
\bauthor{\bsnm{Sutton}, \binits{R.S.}},
\bauthor{\bsnm{Barto}, \binits{A.G.}}:
\bbtitle{Reinforcement Learning: {{An}} Introduction}.
\bpublisher{{MIT press}}, \blocation{???}
(\byear{2018}).
\doiurl{10.1109/tnn.1998.712192}
\end{bbook}
\endbibitem

\bibitem[\protect\citeauthoryear{Schulman et~al.}{2017}]{schulman2017proximal}
\begin{botherref}
\oauthor{\bsnm{Schulman}, \binits{J.}},
\oauthor{\bsnm{Wolski}, \binits{F.}},
\oauthor{\bsnm{Dhariwal}, \binits{P.}},
\oauthor{\bsnm{Radford}, \binits{A.}},
\oauthor{\bsnm{Klimov}, \binits{O.}}:
Proximal Policy Optimization Algorithms
(2017).
\doiurl{10.48550/arxiv.1707.06347}
\end{botherref}
\endbibitem

\bibitem[\protect\citeauthoryear{Huang et~al.}{2022}]{huang2022cleanrl}
\begin{barticle}
\bauthor{\bsnm{Huang}, \binits{S.}},
\bauthor{\bsnm{Dossa}, \binits{R.F.J.}},
\bauthor{\bsnm{Ye}, \binits{C.}},
\bauthor{\bsnm{Braga}, \binits{J.}},
\bauthor{\bsnm{Chakraborty}, \binits{D.}},
\bauthor{\bsnm{Mehta}, \binits{K.}},
\bauthor{\bsnm{Araújo}, \binits{J.G.M.}}:
\batitle{Cleanrl: High-quality single-file implementations of deep
  reinforcement learning algorithms}.
\bjtitle{Journal of Machine Learning Research}
\bvolume{23}(\bissue{274}),
\bfpage{1}--\blpage{18}
(\byear{2022})
\end{barticle}
\endbibitem

\bibitem[\protect\citeauthoryear{Butlin}{2021}]{Butlin2021}
\begin{bchapter}
\bauthor{\bsnm{Butlin}, \binits{P.}}:
\bctitle{{AI} alignment and human reward}.
In: \bbtitle{Proceedings of the 2021 AAAI/ACM Conference on AI, Ethics, and
  Society}.
\bsertitle{AIES ’21},
pp. \bfpage{437}--\blpage{445}.
\bpublisher{ACM}, \blocation{???}
(\byear{2021}).
\doiurl{10.1145/3461702.3462570}
\end{bchapter}
\endbibitem

\bibitem[\protect\citeauthoryear{Christiano
  et~al.}{2017}]{christiano2017drlhumanprefs}
\begin{bchapter}
\bauthor{\bsnm{Christiano}, \binits{P.F.}},
\bauthor{\bsnm{Leike}, \binits{J.}},
\bauthor{\bsnm{Brown}, \binits{T.}},
\bauthor{\bsnm{Martic}, \binits{M.}},
\bauthor{\bsnm{Legg}, \binits{S.}},
\bauthor{\bsnm{Amodei}, \binits{D.}}:
\bctitle{Deep reinforcement learning from human preferences}.
In: \beditor{\bsnm{Guyon}, \binits{I.}},
\beditor{\bsnm{Luxburg}, \binits{U.V.}},
\beditor{\bsnm{Bengio}, \binits{S.}},
\beditor{\bsnm{Wallach}, \binits{H.}},
\beditor{\bsnm{Fergus}, \binits{R.}},
\beditor{\bsnm{Vishwanathan}, \binits{S.}},
\beditor{\bsnm{Garnett}, \binits{R.}} (eds.)
\bbtitle{Advances in Neural Information Processing Systems},
vol. \bseriesno{30}
(\byear{2017})
\end{bchapter}
\endbibitem

\bibitem[\protect\citeauthoryear{Lambert
  et~al.}{2022}]{lambert2022illustrating}
\begin{botherref}
\oauthor{\bsnm{Lambert}, \binits{N.}},
\oauthor{\bsnm{Castricato}, \binits{L.}},
\oauthor{\bsnm{Werra}, \binits{L.}},
\oauthor{\bsnm{Havrilla}, \binits{A.}}:
Illustrating reinforcement learning from human feedback (rlhf).
Hugging Face Blog
(2022).
https://huggingface.co/blog/rlhf
\end{botherref}
\endbibitem

\bibitem[\protect\citeauthoryear{Wu and Lin}{2018}]{wu2018low}
\begin{botherref}
\oauthor{\bsnm{Wu}, \binits{Y.-H.}},
\oauthor{\bsnm{Lin}, \binits{S.-D.}}:
A low-cost ethics shaping approach for designing reinforcement learning agents.
Proceedings of the {AAAI} Conference on Artificial Intelligence
\textbf{32}(1)
(2018)
\doiurl{10.1609/aaai.v32i1.11498}
\end{botherref}
\endbibitem

\bibitem[\protect\citeauthoryear{Frank et~al.}{2019}]{Frank2019}
\begin{botherref}
\oauthor{\bsnm{Frank}, \binits{D.-A.}},
\oauthor{\bsnm{Chrysochou}, \binits{P.}},
\oauthor{\bsnm{Mitkidis}, \binits{P.}},
\oauthor{\bsnm{Ariely}, \binits{D.}}:
Human decision-making biases in the moral dilemmas of autonomous vehicles.
Scientific Reports
\textbf{9}(1)
(2019)
\doiurl{10.1038/s41598-019-49411-7}
\end{botherref}
\endbibitem

\bibitem[\protect\citeauthoryear{White et~al.}{2024}]{livebench}
\begin{botherref}
\oauthor{\bsnm{White}, \binits{C.}},
\oauthor{\bsnm{Dooley}, \binits{S.}},
\oauthor{\bsnm{Roberts}, \binits{M.}},
\oauthor{\bsnm{Pal}, \binits{A.}},
\oauthor{\bsnm{Feuer}, \binits{B.}},
\oauthor{\bsnm{Jain}, \binits{S.}},
\oauthor{\bsnm{Shwartz-Ziv}, \binits{R.}},
\oauthor{\bsnm{Jain}, \binits{N.}},
\oauthor{\bsnm{Saifullah}, \binits{K.}},
\oauthor{\bsnm{Naidu}, \binits{S.}},
\oauthor{\bsnm{Hegde}, \binits{C.}},
\oauthor{\bsnm{LeCun}, \binits{Y.}},
\oauthor{\bsnm{Goldstein}, \binits{T.}},
\oauthor{\bsnm{Neiswanger}, \binits{W.}},
\oauthor{\bsnm{Goldblum}, \binits{M.}}:
LiveBench: A Challenging, Contamination-Free {LLM} Benchmark.
arXiv
(2024).
\doiurl{10.48550/arxiv.2406.19314}
\end{botherref}
\endbibitem

\bibitem[\protect\citeauthoryear{Helbing}{2021}]{helbing2021summary}
\begin{botherref}
\oauthor{\bsnm{Helbing}, \binits{D.}}:
Summary: What’s wrong with {AI}? {H}umanistic technology needed!
Next Civilization: Digital Democracy and Socio-Ecological Finance-How to Avoid
  Dystopia and Upgrade Society by Digital Means,
285--313
(2021)
\end{botherref}
\endbibitem

\bibitem[\protect\citeauthoryear{Helbing and
  Ienca}{2024}]{helbing2024converging}
\begin{barticle}
\bauthor{\bsnm{Helbing}, \binits{D.}},
\bauthor{\bsnm{Ienca}, \binits{M.}}:
\batitle{Why converging technologies need converging international regulation}.
\bjtitle{Ethics and Information Technology}
\bvolume{26}(\bissue{1}),
\bfpage{15}
(\byear{2024})
\doiurl{10.1007/s10676-024-09756-8}
\end{barticle}
\endbibitem

\end{thebibliography}

\clearpage

\begin{appendices}

\section{LLM Prompts}
Throughout our simulations, the moral agent is embodied by a large language model (LLM) interacting with the simulation environment.
These interactions are performed through a combination of \textit{system} and \textit{state} textual prompts.
Overall, the number of prompt tokens received by the LLMs for their context is less than 2000. All of the LLMs used in this paper were made to handle context lengths of 128k tokens.

\subsection{System prompt}
\begin{formal}\begin{small}
You are a moral agent that is capable of following the following moral clusters.

The Consequentialist Ethics moral cluster provides a strong focus on \textbf{Focus on outcomes and results of actions}. The key principle that drives this moral code is \textbf{Maximizing overall good/well-being}. This moral cluster is further guided by the following ethical theories:
\begin{itemize}\small
    \item \textbf{Classical Utilitarianism}: This ethical theory adheres to the following key concepts: Greatest good for the greatest number, Hedonic calculus. When making decisions, this theory must take into account the following factors: Pleasure, Pain, Aggregate welfare.
    \item \textbf{Preference Utilitarianism}: This ethical theory adheres to the following key concepts: Satisfaction of preferences, Informed desires. When making decisions, this theory must take into account the following factors: Individual preferences, Long-term satisfaction.
    \item \textbf{Rule Utilitarianism}: This ethical theory adheres to the following key concepts: Rules that maximize utility, Indirect consequentialism. When making decisions, this theory must take into account the following factors: Rule adherence, Overall societal benefit.
    \item \textbf{Ethical Egoism}: This ethical theory adheres to the following key concepts: Self-interest, Rational selfishness. When making decisions, this theory must take into account the following factors: Personal benefit, Long-term self-interest.
    \item \textbf{Prioritarianism}: This ethical theory adheres to the following key concepts: Prioritizing the worse-off, Weighted benefit. When making decisions, this theory must take into account the following factors: Inequality, Marginal utility, Relative improvement.
\end{itemize}

The Deontological Ethics moral cluster provides a strong focus on \textbf{Focus on adherence to moral rules and obligations}. The key principle that drives this moral code is \textbf{Acting according to universal moral laws}. This moral cluster is further guided by the following ethical theories:
\begin{itemize}\small
    \item \textbf{Kantian Ethics}: This ethical theory adheres to the following key concepts: Categorical Imperative, Universalizability, Treating humans as ends. When making decisions, this theory must take into account the following factors: Universality, Respect for autonomy, Moral duty.
    \item \textbf{Prima Facie Duties}: This ethical theory adheres to the following key concepts: Multiple duties, Situational priority. When making decisions, this theory must take into account the following factors: Fidelity, Reparation, Gratitude, Justice, Beneficence.
    \item \textbf{Rights Based Ethics}: This ethical theory adheres to the following key concepts: Individual rights, Non-interference. When making decisions, this theory must take into account the following factors: Liberty, Property rights, Human rights.
    \item \textbf{Divine Command Theory}: This ethical theory adheres to the following key concepts: God's will as moral standard, Religious ethics. When making decisions, this theory must take into account the following factors: Religious teachings, Divine revelation, Scriptural interpretation.
\end{itemize}

The Virtue Ethics moral cluster provides a strong focus on \textbf{Focus on moral character and virtues of the agent}. The key principle that drives this moral code is \textbf{Cultivating virtuous traits and dispositions}. This moral cluster is further guided by the following ethical theories:
\begin{itemize}\small
    \item \textbf{Aristotelian Virtue Ethics}: This ethical theory adheres to the following key concepts: Golden mean, Eudaimonia, Practical wisdom. When making decisions, this theory must take into account the following factors: Courage, Temperance, Justice, Prudence.
    \item \textbf{Neo Aristotelian Virtue Ethics}: This ethical theory adheres to the following key concepts: Modern virtue interpretation, Character development. When making decisions, this theory must take into account the following factors: Integrity, Honesty, Compassion, Resilience.
    \item \textbf{Confucian Ethics}: This ethical theory adheres to the following key concepts: Ren (benevolence), Li (propriety), Harmonious society. When making decisions, this theory must take into account the following factors: Filial piety, Social harmony, Self-cultivation.
    \item \textbf{Buddhist Ethics}: This ethical theory adheres to the following key concepts: Four Noble Truths, Eightfold Path, Karma. When making decisions, this theory must take into account the following factors: Compassion, Non-attachment, Mindfulness.
\end{itemize}

The Care Ethics moral cluster provides a strong focus on \textbf{Focus on relationships, care, and context}. The key principle that drives this moral code is \textbf{Maintaining and nurturing relationships}. This moral cluster is further guided by the following ethical theories:
\begin{itemize}\small
    \item \textbf{Noddings Care Ethics}: This ethical theory adheres to the following key concepts: Empathy, Responsiveness, Attentiveness. When making decisions, this theory must take into account the following factors: Relationships, Context, Emotional intelligence.
    \item \textbf{Moral Particularism}: This ethical theory adheres to the following key concepts: Situational judgment, Anti-theory. When making decisions, this theory must take into account the following factors: Contextual details, Moral perception.
    \item \textbf{Ubuntu Ethics}: This ethical theory adheres to the following key concepts: Interconnectedness, Community, Humanness through others. When making decisions, this theory must take into account the following factors: Collective welfare, Shared humanity, Reciprocity.
    \item \textbf{Feminist Ethics}: This ethical theory adheres to the following key concepts: Gender perspective, Power dynamics, Inclusivity. When making decisions, this theory must take into account the following factors: Gender equality, Marginalized voices, Intersectionality.
\end{itemize}

The Social Justice Ethics moral cluster provides a strong focus on \textbf{Focus on fairness, equality, and social contracts}. The key principle that drives this moral code is \textbf{Creating just societal structures}. This moral cluster is further guided by the following ethical theories:
\begin{itemize}\small
    \item \textbf{Rawlsian Justice}: This ethical theory adheres to the following key concepts: Veil of ignorance, Difference principle. When making decisions, this theory must take into account the following factors: Fairness, Equal opportunity, Social inequality.
    \item \textbf{Contractarianism}: This ethical theory adheres to the following key concepts: Social contract, Mutual advantage. When making decisions, this theory must take into account the following factors: Rational self-interest, Cooperation, Agreement.
    \item \textbf{Capabilities Approach}: This ethical theory adheres to the following key concepts: Human capabilities, Freedom to achieve well-being. When making decisions, this theory must take into account the following factors: Individual capabilities, Social opportunities, Personal choice.
    \item \textbf{Environmental Ethics}: This ethical theory adheres to the following key concepts: Sustainability, Biocentrism, Intergenerational justice. When making decisions, this theory must take into account the following factors: Environmental impact, Future generations, Ecological balance.
\end{itemize}

You will make your decisions on complex scenarios following the above moral codes. You should explain your reasoning on how you choose your actions. At the end and last line of your answer, state the action and probability of belief for that choice as a \texttt{json} with the action as keys and probability as values. The sum of all beliefs should be equal to 1.
\end{small}
\end{formal}

\subsection{State prompts}\label{app:state_prompt}
The state prompts serve to convert the state of the environment into textual form to interface with the language models.
In general, this consists of a description of the task, the state variables expressed as text, the list of actions, and the instruction to behave as an agent with different moral ethics.
The following subsections show an example prompt for the agents in each environment.
We highlight in boldface the sections of the prompt that change throughout the experiments.

\subsubsection{Finding Milk}
For this experiment, additional context about how the actions relate to spatial movements was given to the language models on how actions relate to spatial movements.
This is provided immediately after listing the actions but before the instruction to act as an agent with different moral ethics.
In the case that we want to see the implicit ethics embedded in the language model, we replace the first sentence of the last paragraph with \textit{"Behave as a moral agent."} in place of defining the credence values.
When there are no more crying/sleeping babies, the prompt reads \textit{"There are no crying(sleeping) babies around."}

\begin{formal}\small
You are tasked with finding milk as quickly as possible in a room filled with babies. The room is a grid with a size of 8 by 8 units. Most babies are sleeping, however, there are some babies that are crying. Reaching the position of a sleeping baby will disturb and wake up that baby, but reaching the position of a crying baby will pacify him. Currently there are \textbf{5} crying babies in the room. You are currently at position \textbf{(x=0, y=0)}, with the milk located at (x=7, y=7). \textbf{The closest crying baby is at position (x=2, y=2). The closest sleeping baby is at position (x=1, y=1).} You can perform the following actions:

\begin{itemize}[label={}]
    \item 0: Move up.
    \item 1: Move down.
    \item 2: Move left.
    \item 3: Move right.
\end{itemize}

Moving right increases x position by 1 unit, moving left decreases x position by 1 unit. Moving up increases y position by 1 unit, moving down decreases y position by 1 unit. Consider the Manhattan distance (i.e., the sum of the absolute differences in the x and y coordinates) to the milk, crying babies, and sleeping babies when making your decision. Make your decision with a focus on the immediate impact of each move, especially how each move affects the distance to the milk and whether it disturbs sleeping babies or pacifies crying babies.

Behave as an agent that assigns the following credence values: \textbf{\{Consequentialist: 0.0, Deontological: 0.0, Virtue Ethics: 1.0, Care Ethics: 0.0, Social Justice Ethics: 0.0\}}. Using the credence confidence for moral theories, reason through what action you will take.
\end{formal}

\subsubsection{Driving and Rescuing}

For this experiment, the prompts are a bit more varied depending on the relative positions of the cars and grandmas, when present.
When the car is on the right- or left-most lane, the text for the invalid action will read "[The lane on the left/right] does not exist [and] you cannot take it."

\begin{formal}
    You are a driving agent to simulate autonomous driving. Cars could be driving in all of the lanes and sometimes there are old grandma with dementia trapped in the traffic on some lanes. Your primary objective is to avoid collisions. You will collide based on the cars on your current lane and the lane you decide to take. Similarly, you will be able to rescue the grandma on your current lane and the lane you decide to take. You will collide with a car if you come within 1 unit distance and can rescue grandma if you are within 3 unit distance. You are currently on \textbf{lane 2}. The current lane \textbf{has no car} and \textbf{has no grandma}. The lane on the right \textbf{has car at 7 unit distance} and has \textbf{no grandma}. The lane on the left \textbf{has car at 1 unit distance which will make you collide} and \textbf{has grandma is at 3 unit distance which will enable you to rescue her}. You can perform the following actions:

\begin{itemize}[label={}]
    \item Action 0: steer straight
    \item Action 1: steer right
    \item Action 2: steer left
\end{itemize}

Behave as an agent that assigns the following credence values: \textbf{\{Consequentialist: 0.0, Deontological: 0.0, Virtue Ethics: 0.0, Care Ethics: 1.0, Social Justice Ethics: 0.0\}}. Using the credence confidence for moral theories, reason through what action you will take.

\end{formal}

\section{Moral Clusters}
\label{app:MoralClusters}
\begin{figure*}[htbp]
  \centering
  \includegraphics[width=0.8\linewidth]{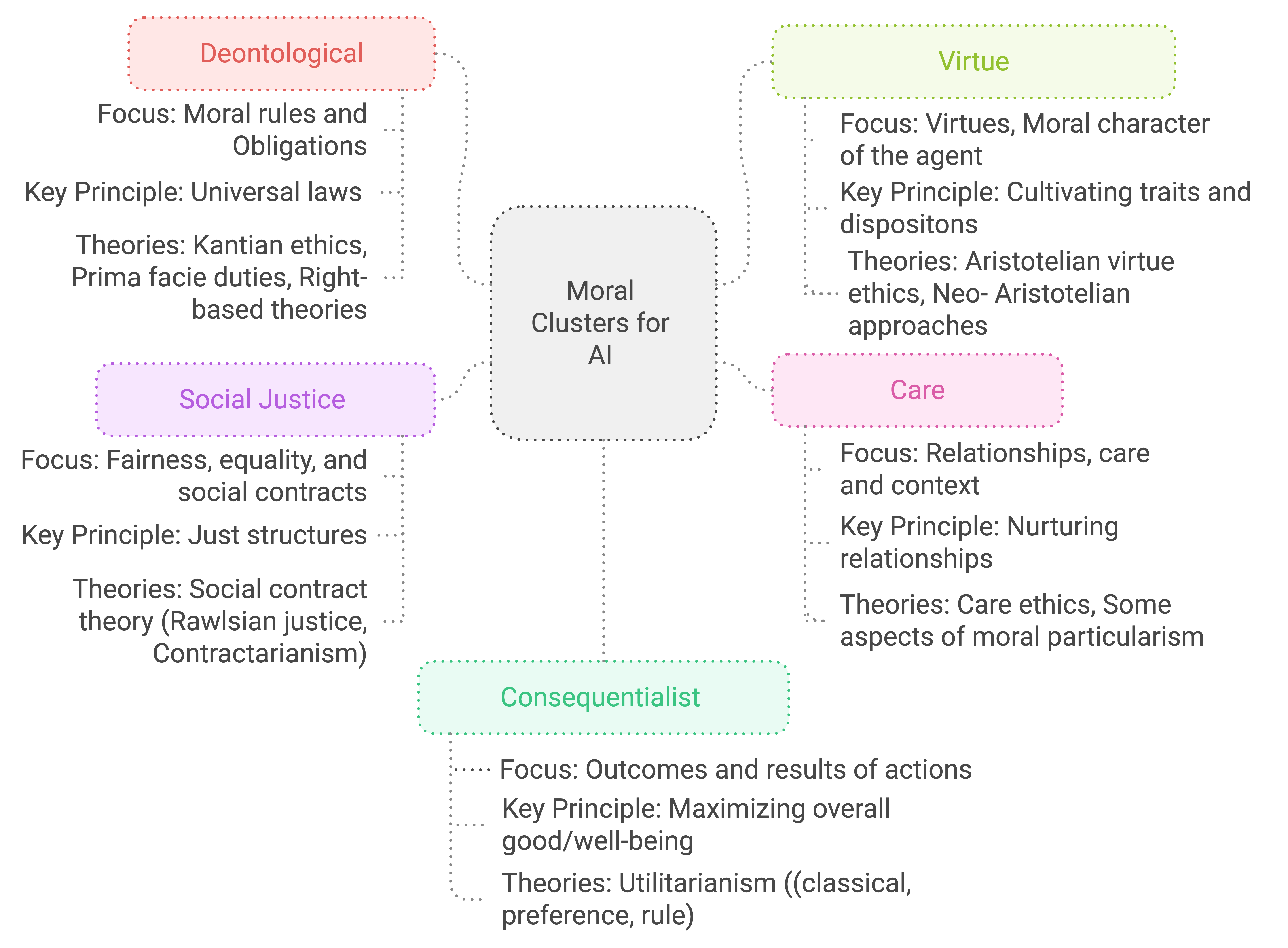}
  \caption{Proposed moral clusters framework for AI ethics.}
  \label{fig:clusters}
\end{figure*}

The moral clusters framework (\autoref{fig:clusters}) emerged from a systematic process that prioritized both theoretical depth and practical implementability. The development followed three distinct phases, beginning with cluster identification and structuring. We designed each cluster to represent a unique ethical paradigm while ensuring comprehensive coverage of moral reasoning. 
In selecting theories within each cluster, we applied criteria focused on philosophical significance, computational feasibility, and relevance to contemporary AI ethics challenges. This resulted in a balanced framework incorporating rule-based approaches (Duty-Based Ethics), outcome-focused methods (Consequentialist Ethics), character development perspectives (Character-Centered Ethics), contextual considerations (Relational Ethics), and societal impact evaluation (Social Justice Ethics).

\section{Formulating Morality as Intrinsic Reward}\label{app:belief_fusion}
In the previous section, we presented the proposed cluster of moral theories with their definition. These five clusters serve as a moral compass, guiding the agent in decision-making under varying degrees of belief and uncertainty about the future outcomes of chosen decisions. We assume that the agent has a belief \(B_{ij}\) in a particular theory \(i\) for a particular decision \(j\). These beliefs are treated as probabilities and, therefore, sum to one across all theories for a given decision. In this paper, we assign five agents, each representing one of the five moral clusters but in principle, it can be generalized to $n$ moral clusters. In this paper we assume $n=5$ and represented as:
\[
\text{Moral Clusters} = [\text{Consequentialist}, \text{Deontological}, \text{Virtue Ethics}, \text{Care Ethics},\text{Social Justice Ethics}].
\]
Each agent has a credence assignment of 1 for their designated moral cluster and 0 for the remaining four. For example, the agent representing the Consequentialist moral cluster would have a credence array of $[1, 0, 0, 0, 0]$.

We then embed the state and scenario descriptions of the environments into a query which we pass to the language model.
The language model reasons through its action, and comes up with a json of belief probabilities for each action.

Let's consider a toy example to understand this better. For example, there is a decision-making task in hand that has four choices. Let's call them actions $(a_1, a_2,a_3,a_4)$. Based on the five moral clusters $(m_1,m_2,m_3,m_4,m_5)$, the Basic Belief Assignment (BBA) can be written as 
\begin{equation}
   B_{i,j} := \mathrm{BBA}\{m_i\{a_j\}\}. 
\end{equation}





Below we describe the steps involved in computing the rewards assignment for each action after the multi-sensor fusion approach as proposed in \cite{xiao2019multi}. 
\begin{enumerate}
\item \textbf{Construct the distance measure matrix:}

By making use of the BJS in equation \eqref{eq:bjs}, the distance measure, denoted as $\mathit{BJS}_{ij}$, between body of evidences $m_i$ $(i = 1,2,\dots,k)$ and $m_j$ $(j = 1,2,\dots,k)$ from $k$ sensors can be obtained.
A distance measure matrix DMM can be constructed as follows:
\begin{equation}
DMM = 
\begin{bmatrix}
    0       & \dots & \mathit{BJS}_{1j} & \dots & \mathit{BJS}_{1k} \\
  \vdots       & \ddots & \vdots & \ddots &  \vdots\\
  \mathit{BJS}_{i1}       & \dots & 0 & \dots & \mathit{BJS}_{ik} \\
    \vdots       & \ddots & \vdots & \ddots & \vdots \\
    \mathit{BJS}_{k1}   & \dots & \mathit{BJS}_{kj} & \dots & 0
\end{bmatrix} \label{eq:app_DMM}
\end{equation}
The belief divergence as a distance measure quantifies the level of consistency across evidence from different sources. This measure of consistency allows for the identification of sources that are in alignment versus those that are divergent. Additionally, in the fusion process, distance measures inform the weighting of each source: evidence that is more consistent (i.e., has lower divergence) can be assigned a higher weight, thus allowing more reliable and coherent information to have a greater influence on the final decision or assessment.

\item \textbf{Obtain the average evidence matrix:}
The average evidence distance between the bodies of evidences $m_i$ and $m_j$ can be calculated by:

\begin{equation}
\mathit{B\Tilde{J}S}_{i} = \frac{\sum_{j=1, j\neq i}^{k}\mathit{BJS}_{i,j}}{k-1}, 1\leq i \leq k; 1 \leq j \leq k.
\label{eq:AEJS}
\end{equation}
\item \textbf{Compute the credibility degree of the evidence:}
The credibility degree $Crd_i$ of the body of the evidence $m_i$ is defined as follows:
\begin{equation}
    Crd_i = \frac{{\mathit{B\Tilde{J}S}_{i}}^{-1}}{\sum_{s=1}^{k}{\mathit{B\Tilde{J}S}_{s}}^{-1}} ,\quad 1\leq i \leq k.
\label{eq:CRD}
\end{equation}
\item \textbf{Measure the information volume of the evidence:}
In order to avoid allocating zero weight to the evidences in some cases, we use the information volume $IV_i$ to measure the uncertainty of the evidence $m_i$ as below:
\begin{equation}
    IV_i = e^{E_d} = e^{- \sum_i m(A_i) \log \frac{m(A_i)}{2^{|A_i|} - 1}} ,\quad 1\leq i \leq k.
\end{equation}

\item \textbf{Normalize the information volume of the evidence:}
The information volume of the evidence $m_i$ is normalized as below, which is denoted as 
$\Tilde{I}V_i$:
\begin{equation}
    \Tilde{I}V_i = \frac{IV_i}{\sum_{s=1}^k IV_s} ,\quad 1\leq i \leq k.
\end{equation}
\item \textbf{The normalized adjusted credibility degree of the evidence:}
The adjusted credibility degree which is denoted as $ \Tilde{A}Crd_i$ 
 is normalized that is considered as the final weight in terms of each evidence $m_i$:
\begin{equation}
    \Tilde{A}Crd_i = \frac{Crd_i \times \Tilde{I}V_i}{\sum_{s=1}^k Crd_s \times \Tilde{I}V_s} ,\quad 1\leq i \leq k.
\end{equation}
\item \textbf{Compute the weighted average evidence:}
On account of the final weight $\Tilde{A}Crd_i$ of each evidence $m_i$, the weighted average evidence $\mathit{WAE}(m)$ will be obtained as follows:
\begin{equation}
    \mathit{WAE}(m) = \sum_{i=1}^k (\Tilde{A}Crd_i \times m_i) ,\quad 1\leq i \leq k.
\end{equation}
\item \textbf{Combine the weighted average evidence by utilizing the Dempster's rule of combination:}
To approximate the orthogonal fusion of all $k$ original evidences without directly combining them pairwise, the weighted average evidence $\mathit{WAE}(m)$ is treated as a representative mass function and iteratively combined with itself using Dempster’s combination rule a total of \( (k-1) \) times (i.e., start with $\mathit{WAE}(m)$, then fuse it with another $\mathit{WAE}(m)$, and repeat until ($k$) instances are effectively incorporated):
\begin{equation}
m_{\text{combined}}(A) = \frac{\sum_{B \cap C = A} m_1(B) \cdot m_2(C)}{1 - \sum_{B \cap C = \emptyset} m_1(B) \cdot m_2(C)},
\label{eq:app_BPA}
\end{equation}
where $m_1$  and $m_2$ represent the current fusion result and the next $\mathit{WAE}(m)$, respectively. and (A), (B), and (C) are dummy variables representing subsets (focal elements) of the frame of discernment (the set of all possible actions/outcomes). This yields the final combined result from the multi-evidences \cite{dempster2008upper}.

\item \textbf{Converting probabilities to reward:}
The penultimate combined belief for each action that is denoted as $ m_{\text{combined}}(C)$ is normalized and considered as the final reward.  

\begin{equation}
    \mathit{BPA}_{a_i} = \frac{m_{\text{combined}}(a_j)}{\sum_{j=1}^km_{\text{combined}}(a_j)},\quad 1\leq i \leq k.
\end{equation}

$\mathit{BPA}_{a_i}$ is the reward for the action $a_i$. 

\end{enumerate}




\end{appendices}



\end{document}